**Title**

# The role of evolutionary selection in the dynamics of protein structure evolution

**Condensed Title**

The pace of protein structure evolution

**Authors**

Amy I. Gilson[1], Ahmee Marshall-Christensen[1], Jeong-Mo Choi[1] & Eugene I. Shakhnovich[1]

**Affiliations**

[1]Department of Chemistry and Chemical Biology, Harvard University, 12 Oxford St., Cambridge MA, USA

**Corresponding author**

Eugene I. Shakhnovich, Department of Chemistry and Chemical Biology, Harvard University, 12 Oxford St., Cambridge MA, USA; e-mail: shakhnovich@chemistry.harvard.edu

**Abstract**
Homology modeling is a powerful tool for predicting a protein's structure. This approach is successful because proteins whose sequences are only 30% identical still adopt the same structure, while structure similarity rapidly deteriorates beyond the 30% threshold. By studying the divergence of protein structure as sequence evolves in real proteins and in evolutionary simulations, we show that this non-linear sequence-structure relationship emerges as a result of selection for protein folding stability in divergent evolution. Fitness constraints prevent the emergence of unstable protein evolutionary intermediates thereby enforcing evolutionary paths that preserve protein structure despite broad sequence divergence. However on longer time scales, evolution is punctuated by rare events where the fitness barriers obstructing structure evolution are overcome and discovery of new structures occurs. We outline biophysical and evolutionary rationale for broad variation in protein family sizes, prevalence of compact structures among ancient proteins and more rapid structure evolution of proteins with lower packing density.

## Introduction

A wide variety of protein structures exist in nature, however the evolutionary origins of this panoply of proteins remain unknown. While protein sequence evolution is easily traced in nature and produced in the laboratory, the emergence of new protein structures is rarely observed and difficult to engineer (1-3). One approach to studying structure evolution is to examine how proteins' structural similarity varies over a range of sequence identities. Such investigations proceed by aligning many pairs of proteins so that their sequence identity (or another measure of sequence similarity) and structural similarity can be assessed (4-8). The result is a cusped relationship between sequence and structure divergence: sequences reliably diverge up to 70% without significant protein structure evolution. Below 30% sequence identity, the structural similarity between proteins abruptly decreases, giving rise to a "twilight zone" where little can be said about the relationship between sequence identity and structural similarity without more advanced methods. This finding is the foundation of one of the most important methods in protein biophysics: structure homology modeling (9, 10). Despite the fact that the plateau of high structural similarity above 30% sequence identity has been crucial for homology modeling and that many of the advanced structure prediction methods have been motivated by abrupt onset of the twilight zone, the cusped relationship between sequence and structural similarity has not yet received a detailed biophysical justification (11, 12).

Previous work characterized the relationship between sequence and structure similarity by fitting the data empirically with an exponential function, and the adequacy of this model was interpreted as evidence in favor of the local model of protein structure determination, namely, that only a key subset of residues encode a protein's structure (4, 5, 7). However we are not aware of any further evidence that mutating a special subset, amounting to about 30% of a protein's residues, generally causes a protein's structure to evolve to a new structure. Conversely, randomly mutating 70% of a protein's residues will almost surely unfold it as even a small number of point mutations can destroy a protein's structure (13). Therefore, it is clear that without evolutionary selection, the range of 100-30% sequence identity could not correspond to nearly identical structures.

Purely physical models of structure evolution, without any selection, have explained many fundamental features of the protein universe. Dokholyan et al. constructed a protein domain universe graph in which protein domain nodes are connected by an edge if they are structurally similar. The resulting graph is scale-free, which they showed would be the result of duplication and structural divergence of proteins (14, 15). Similarly, the birth, death, innovation models developed by Koonin et al. explain the power law-like distribution of gene family sizes that exists in many genomes (16). However, because these works use neutral models, they are unable to explain the cusped sequence-structure relationship.

A small but growing collection of cases where protein structure evolution has been observed or inferred provides mechanistic insight into the role of selection in protein structure evolution. They show that it is possible for proteins to be within a small number of point mutations of a fold evolution event (17-19) but also that structure emergence may often pass through thermodynamically destabilized intermediates. Among Cro bacteriophage transcription factors, a pair of homologous proteins with 40% sequence identity (indicating that they share a recent common ancestor) but different structures was found. Subsequent studies showed that some Cro proteins might be just a few mutations away from changing fold (17, 20). Similarly, a trajectory of point mutations was engineered to convert a natural protein adopting a $3\,\alpha$ fold into a $4\,\beta + \alpha$ fold. Finding a rare sequence of mutations that avoided unfolded evolutionary intermediates was a major achievement of this work (18, 21, 22). Computational investigations of protein structure evolution using model proteins of Protein Database structures also show that structure evolution traverses unstable intermediates and found that less stable proteins are more evolvable at the structural level (23, 24).

Here, we explore the evolutionary dynamics of protein structure discovery. Given the fact that most globular proteins must adopt a well-defined 3D structure in order to function and the strong evidence that many fold evolution pathways require protein destabilization, we hypothesize that protein structure discovery requires crossing valleys of low fitness on fitness landscapes, corresponding to genetic encoding of evolutionarily transient, unstable proteins. Therefore, the strength of selection for folding stability under which a protein evolves may modulate its capacity to evolve a new structure.

We study evolution of protein structures using the data on sequence-structure relationships in natural proteins and a variety of evolutionary models of increasing complexity and realism. We find that the cusped relationship between structure and sequence divergence is a direct consequence of the interplay between evolutionary dynamics and the biophysical constraint for protein folding stability. In both the bioinformatics data and simulation data, the sharpness of the "cusp," but not its position (approximately 30% sequence identity for natural proteins) is determined by the compactness of evolving proteins, a proxy for their thermodynamic stability. Rather than fitting these data empirically, we formalized the mechanism underlying structure-sequence divergence in an "ab-initio" analytical model that fits the bioinformatics data for proteins grouped by their compactness. Simulations show that what underlies the negative correlation between protein stability and structure evolution rate is the strength of evolutionary selection for stability under which proteins evolve. Fitness barriers are imposed by selection for thermodynamic stability and continual sequence evolution degrades sequence identity over the time scales needed to find mutations that overcome these barriers and which encode alternative, stable protein structures. Many long-standing observations in protein biophysics are reinterpreted and unified using this powerful yet simple interpretive framework.

## Results

### Structure-sequence relationship in the protein universe

A cusp-like relationship between structure and sequence divergence has long been observed in the record (4-6, 11) when the sequence identity and structural similarity is determined for many pairs of protein domains. Here, we explore how selection for protein stability affects the shape of the relationship between sequence and structure divergence. Following Chothia and Lesk (4), who studied the divergence of proteins within protein families, we studied the divergence of globular protein domains classified into the same SCOP fold. By choosing a broader classification than family, we could track divergence over the long timescales required for significant structure evolution to be detected (See *Methods*, Fig. 1A) (25, 26). SCOP folds are broadly defined by major secondary structures in similar arrangements and topologies, so protein domains sharing a fold classification often adopt significantly different structures (25). Sequences were aligned using the Needleman-Wunsch (NW) algorithm as implemented in MATLAB and sequence identity was calculated by the number of identical residues matched in the alignment normalized by the average length of the aligned proteins. Structures were aligned using the Template-Modeling (TM) algorithm and given a TM score between 0 and 1 to quantify their degree of structural similarity (27).

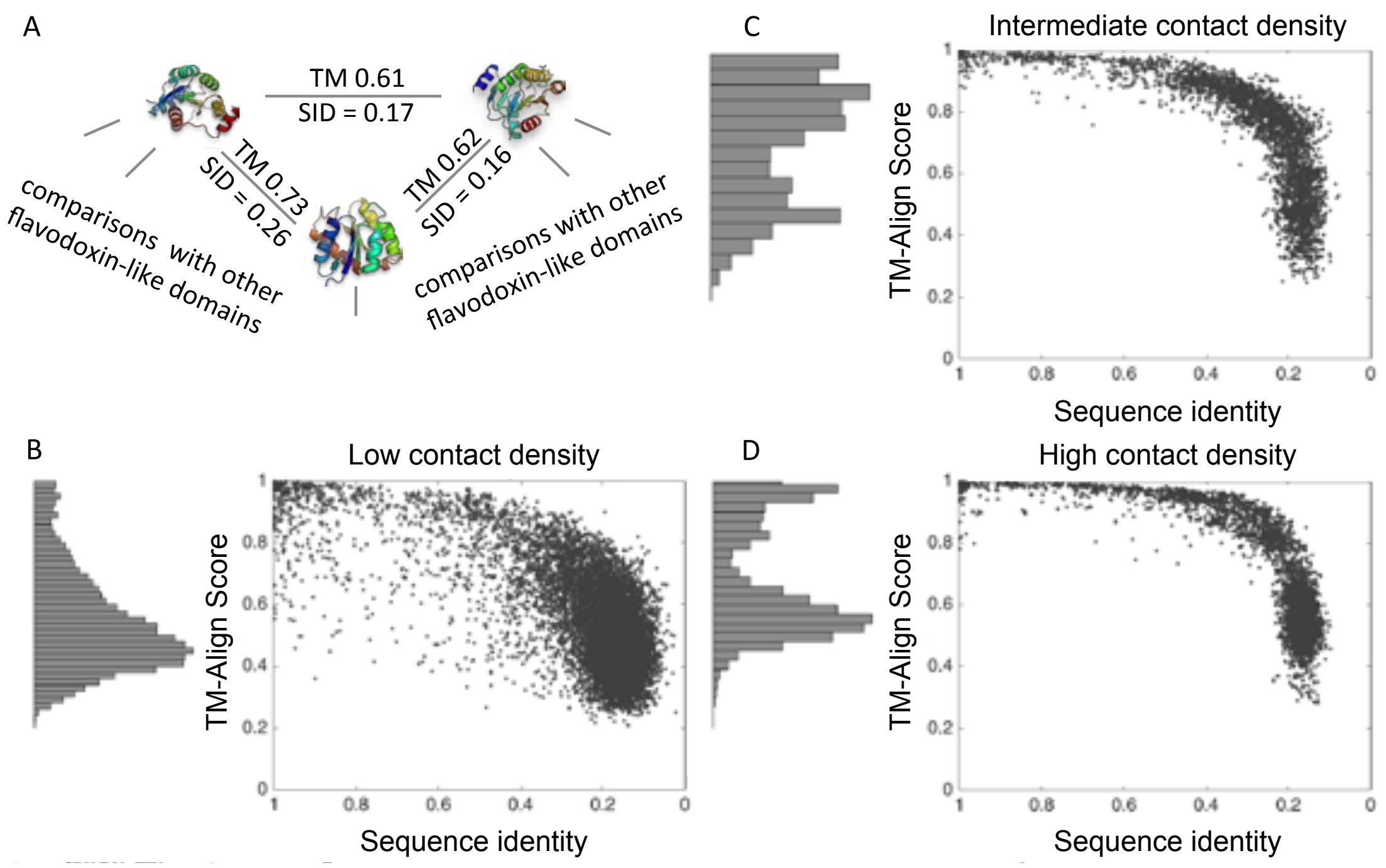

**Figure 1: Stability and fold evolution of SCOP domains** SCOP domains classified as α, β, α+β, α/β were used. Sequence identity (SID) and structural homology (TM-align score) for each pair of domains classified into the same fold were calculated. (A) An illustrative example from the α+β flavodoxin-like folds. The SCOP domains and their associated PDB ids from left to right are d1a04a2 (1A04), d1a2oa1 (1A2O), and d1eudb1 (1EUD). (B-D) The relationship between sequence divergence and structure evolution in SCOP domains. Note that sequence identity decreases from left to right. Domains are partitioned by contact density (CD) (B) bottom 10% contact density (<4.13 contacts/residue, N=12,671 data points) (C) middle 10% contact density (4.57<CD<4.65 contacts/residue. N=3,863 data points) (D) top 10% contact density (>4.93 contacts/residue, N=5,672 data points). Histograms at the left of the plots show, in B, that at low contact density, the distribution of TM-align scores is approximately single peaked while C and D show that at higher contact densities, the distribution of structure similarity TM-align scores is bimodal.

We define thermodynamic stability of a protein as the fraction of molecules in the ensemble that resides in the native state. It is an experimentally observable thermodynamic property of proteins that is related to fitness and as such might be under evolutionary selection (28-33). For single domain proteins that fold as two state systems, this quantity is directly related to the folding free energy $\Delta G$ through the Boltzmann relation of statistical mechanics:

$$P_{nat} = \frac{e^{-\frac{\Delta G}{k_B T}}}{1 + e^{-\frac{\Delta G}{k_B T}}} \qquad (1)$$

where $T$ is temperature and $k_B$ is Boltzmann constant. Here we consider only single domains as defined in SCOP and use the number of amino acid contacts normalized by the domain length, the contact density (CD), as a proxy for the folding free energy $\Delta G$ of a protein (23). This approach compensates for the relative deficiency of experimental thermodynamic data. The connection between CD and folding free energy has been established through several lines of evidence. Inter-residue contacts stabilize a protein structure through van der Waals interactions, so the more contacts in a protein structure the more stable the structure. Indeed, proteins of thermophilic organisms have greater contact density than their mesophilic homologs (34). Finally, protein length correlates with both contact density and folding free energy. Correlation between length and contact density is a consequence of the globular structure of proteins where surface to volume ratio depends on total length. Folding free energy is an extensive property that increases with the total number of amino acids in a protein (35, 36). Generally lower free energy does not necessarily mean greater stability ($P_{nat}$). That is, a protein consisting of two identical thermodynamically independent domains each having folding free energy $\Delta G$, will not be more stable than each domain in separation. However, for single-domain proteins that fold in a two-state manner folding free energy is directly related to stability as can be seen from Eq.1.

In Figs 1B-D (Dataset S1), the data are subdivided by contact density, such that only proteins in the bottom 10% (CD < 4.13 contacts per amino acid residue), middle 10% (4.65 > CD > 4.57 contacts per amino acid residue), and top 10% (CD > 4.93 contacts per amino acid residue), of contact density are compared (Fig 1B, C, and D respectively). For the most stable proteins, there is a bimodal distribution of TM-align scores with a peak at high TM-align scores corresponding to diverging proteins maintaining near identical structures and a peak at low TM-align scores corresponding to proteins that are unrelated or whose folds have evolved (Fig. 1B). However in the class of low contact density proteins there is only a peak at low TM-align score. Furthermore, for high and intermediate contact density domain, there is a pronounced cusp in the sequence-structure relationship. By contrast, the transition from homologous structures to diverged structures is more gradual for the low contact density domains, without a visible cusp. These results are robust to the exact sequence similarity metric used as shown in Fig. S1. There, sequence similarity is measured by the NW alignment score using the BLOSOM30 substitution matrix. Because this score takes similarity between aligned residues into account, e.g. a matched alanine and glycine would lead to a higher sequence similarity score than alanine and arginine, and because it penalizes alignment gaps, the NW alignment score attains greater sensitivity than sequence identity. Thus, this analysis robustly indicates that, as proteins diverge, the onset of structure evolution occurs earlier for less stable proteins than for more stable ones.

A possible confounding factor could be that proteins of different contact densities might be enriched in different structural classes so that our finding presented in Fig. 1 reflect differences in evolutionary scenarios for different protein structural classes. However we can eliminate this possibility because we found that there are no significant differences in the shape of the sequence-structure relationship when the data are subdivided by protein class (Fig. S2).

**A simple analytical model for the twilight zone**

An evolutionary trajectory of sequential amino acid substitutions can be imagined connecting proteins of any two structures. There are many experimental studies supporting the vision of an evolutionary landscape where sequences folding into stable structures are connected by sequences that do not adopt stable structures (21, 33, 37-39). In general, a protein's stability determines the cellular amount of folded, and therefore functional, protein as well as the amount of unfolded protein, which is not only non-functional but also can form toxic aggregates (40). The fitness of an organism is thereby directly related to the ability of proteins to carry out their functions and therefore their stability.

Based on these understandings, we construct a simple model of protein structure evolution on a fitness landscape where sequences encoding high-fitness (thermodynamically stable) proteins form peaks separated by fitness valleys. (Fig. 2A). The model is based on three postulates. First, new structures are discovered in divergent evolution from existing structures. Second, in analogy with chemical kinetics we treat the events leading to structure evolution as an activated process, where wait times between fitness valley crossing events are exponentially distributed. The evolutionary reaction coordinate is a mutational path connecting two protein structure states. A free energy barrier separating two states in chemical kinetics is analogous to the evolutionary barrier comprising sequences that encode unstable proteins separating two stable protein structures that confer high fitness on their carrier organisms. In this model, organisms encoding thermodynamically stable proteins are "fit" because we assume proteome stability is a main component of organismal fitness throughout this paper (29). Third, we postulate that crossing the fitness valley leads to the discovery of novel folds that are structurally dissimilar to the original fold.

We denote $k$ as the rate of structure evolution. Modeling structure evolution as an activated process, the probability, $q(t)$, that an ancestral structure (the structure at time $t=0$) is unchanged at time $t$ follows immediately:

$$q(t)=e^{-kt} \tag{2}$$

Next, it is necessary to substitute the variable of time for the variable of sequence identity for two reasons. First, sequences can reach mutation saturation. Second, this transformation will permit direct comparison between the analytical and the bioinformatics results. The relationship for the expected Hamming distance between the evolving protein at time $t$ with respect to the ancestral protein at time $t=0$, normalized by protein length, is given by

$$S(t)=\frac{l(a-1)}{a}\left(1-\left(1-\frac{a}{l(a-1)}\right)^{t}\right) \tag{2}$$

where $l$ is the length of the protein and $a$ is the number of amino acid types, typically 20 (see Supplementary Text for derivation). This expression only takes into account the point mutations that become fixed in the evolving population because the model only takes into account mutations that might contribute to the emergence of a new structure.

Excluding the time variable from Eqs. 3 and 4, we determine the probability $q$ that a structure has persisted once its sequence is at Hamming distance $S$ from its ancestral $(t=0)$ sequence.

$$q(S)=\exp\left(-k\frac{\log\left(1-S\frac{a}{l(a-1)}\right)}{\log\left(1-\frac{a}{l(a-1)}\right)}\right) \tag{3}$$

Figure 2B shows the dependence of structure survival probability as a function of sequence identity ($q(S)$ v. $1-S$). When $k\approx 1$ the protein is free to diffuse through structure space as easily as it does through sequence space. As proteins become more stable, the barriers between protein structures begin to retard structure evolution giving rise to activated dynamics. When $k \ll 1$, the sequence identity degrades to random before the first structure evolution events take place, resulting in an abrupt decrease in $q(S)$ at very low $S$.

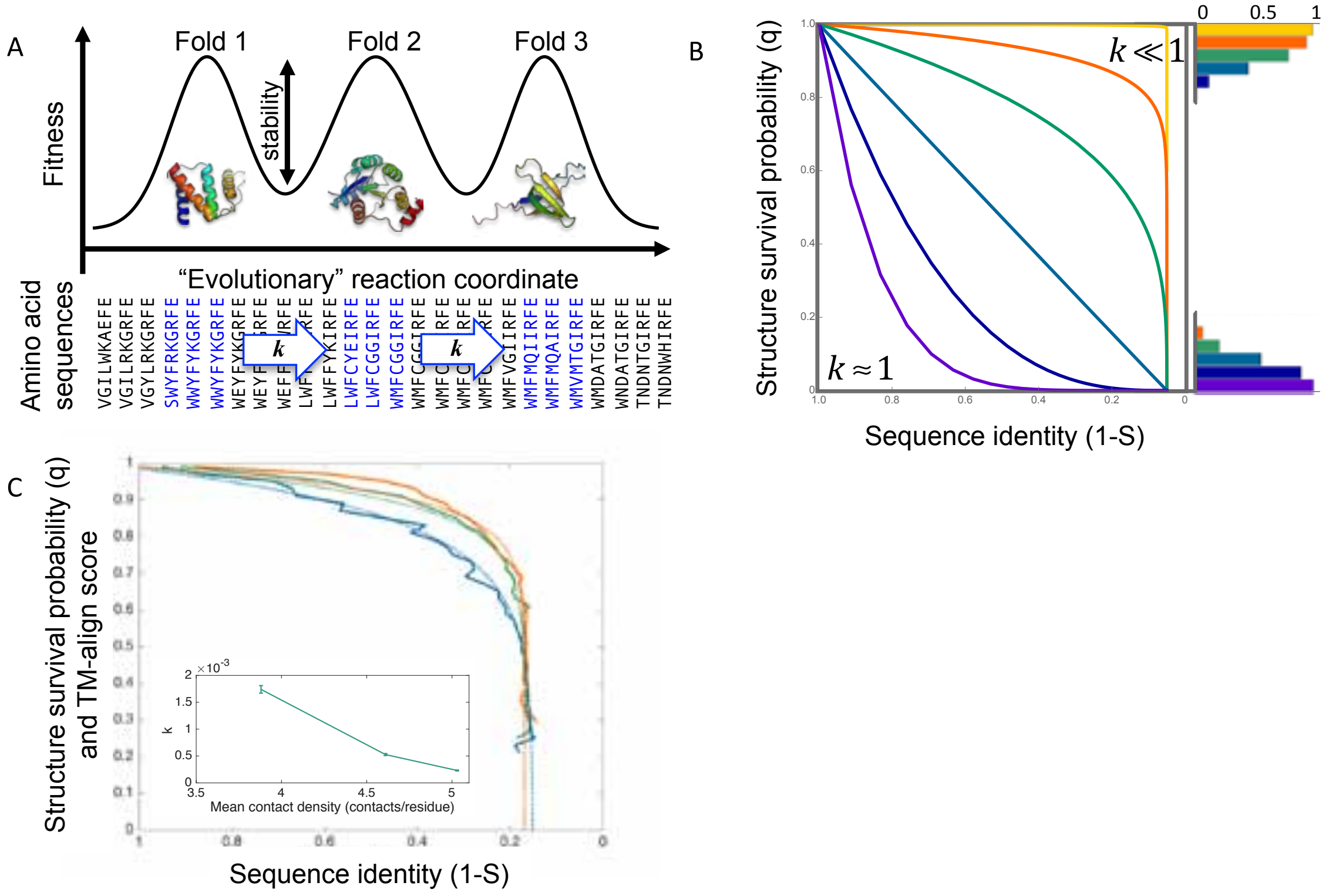

**Figure 2: Fitness barriers to fold evolution leads to cusped sequence-structure relationship** (A) A schematic representation of the evolutionary process in which fold discovery events occur by overcoming fitness valleys along the "evolutionary" reaction coordinate. Sequences that are "more fit," i.e. that contribute to high fitness in the organisms they are part of are shown in blue. Families of homologous proteins are formed by sequence evolution between structure evolution events. The rate of structure discovery, $k$, depends on the thermodynamics stability (a property of protein biophysics) of an evolving proteins because in our model's analogy to chemical kinetics, the more stable an evolving protein, the higher the fitness barrier to evolving a new fold. (B) The relationship between sequence identity and structural survival probability of an ancestral structure for a proteins with length 27 (the length of model proteins used below) at several values of $k$: from top to bottom, 0.0001 (yellow), 0.003 (orange), 0.01 (green), 0.04 (light blue), 0.1 (dark blue) 0.25 (purple). The histograms show the probability that the ancestral structure has persisted (top) versus the probability that new structure has emerged (bottom) after at most 74% of the residues have been mutated. (C) Fit of the analytical model to real protein data. The model fit to the data is presented in dashed lines while binned bioinformatics data is presented in solid lines for each contact density subgroup: low contact density (blue), intermediate contact density (green), high contact density (orange). In the inset, the correlation between protein domain contact density and protein structure evolutionary rate, $k$, predicted analytically. Error bars indicate one standard deviation.

Conceptually, this is because broad exploration of sequence space takes place over the course of many mutations before there arises a sequence that is stable yet which has a new minimum free energy native state. The histograms in Figure 2B also indicate that differences in stability may explain heterogeneity that has been observed in gene family sizes, defined as the number of non-redundant sequences that adopt highly homologous structures (41). For each of the structure evolution rates tested, the histogram shows the probability that the structure has (bottom) or has not (top) evolved after some evolutionary time. The curves and histograms for slow structure evolution rates ($k \ll 1$) are consistent with the high contact density class of proteins while those with intermediate values of $k$ are consistent with the low contact density class of proteins. Unstable proteins move rapidly through structure space, providing little time for gene duplication and sequence divergence to populate a particular family when compared to their thermodynamically stable counterparts.

Finally, we confirm that the analytical model predicts a decreasing rate of structure evolution, $k$, when fit to the three protein domain subgroups with increasing contact density. The bioinformatics data featured in Fig. 1 was binned and averaged for each contact density subgroup and the rate of structure evolution, $k$, and twilight zone offset, $c$, were then fit to this bioinformatics data (See *Methods*). As shown in Figure 2C, this fit confirms both that the analytical model correctly reproduces the shape of the sequence-structure relation for proteins of different contact densities, and that contact density correlates negatively with protein structure evolution rate, $k$, (inset, Table S1). Contact density also correlates negatively with $k$ when the NW alignment score is used to measure sequence similarity (Fig. S3, Table S2). Interestingly, while it might be expected that the analytical model reproduces the correlation between contact density and $k$ when comparing protein domain groups with very different contact densities, it also seems to be able to discriminate between protein subgroups with small differences in average contact density. Protein domains in the four structural classes, $\alpha$, $\beta$, $\alpha/\beta$, $\alpha+\beta$, have average contact densities that range from 4.33 contacts/residue ($\alpha$) to 4.66 contacts/residue ($\alpha/\beta$). While differences in the shape of the sequence-structure relation among the four classes remain hardly distinguishable by eye even after this binning and fitting, the fitted $k$ values do decrease monotonically with the classes' increasing contact density indicating that structure evolution rates are sensitive even to modest differences in contact density (Fig. S4, Table S1).

**Structure evolution of model proteins**

We now turn to explicit modeling of protein structure evolution in order to test the assumptions of our analytical model and to get mechanistic insights into the biophysics of fold emergence. Details of the model are provided in the *Methods* section. In brief, each model protein consists of 27 amino acid residues that fold into a compact $3\times3\times3$ cube (42-44). All 103,346 possible compact structures of such model proteins have been enumerated, and, following Heo et al., we use a representative subset of randomly selected 10,000 conformations as our space of possible protein structures for computational efficiency throughout this work (45). Neighboring amino acid residues that are not connected by a covalent bond interact according to a Miyazawa-Jernigan (MJ) potential (46). In line with previous discussion, the 'fitness' of each each model protein is represented by its stability, $P_{nat}$, the Boltzmann probability that a protein adopts its lowest energy (native) state. For any 27-mer sequence, $P_{nat}$ can be determined exactly within this model (see *Methods*).

Hypothesizing that the strength of evolutionary selection under which a protein evolves is the origin of both the protein's stability and its structure evolution rate, we ran many evolutionary simulations where proteins could evolve new structures under stability ($P_{nat}$) constraints of various stringencies. Each simulation started with a stable protein ($P_{nat} > 0.99$) and each generation, an amino acid substitution was introduced into the evolving protein. Stabilizing mutations that increase $P_{nat}$ were always accepted, while destabilizing mutations were accepted according to the Metropolis criterion with a selective temperature $T_{sel}$ that establishes stringency of evolutionary selection (47, 48) (see *Methods*, Fig. S5 and Dataset S2). Simulations ran for 1,000 generations (mutation attempts, illustrated in Fig. 3A) and the structure, sequence, and stability ($P_{nat}$) were recorded every 10 generations.

First, we quantify the structural similarity between the "wild-type" and mutant structure for each time step where a structure discovery event occurs in order to test the structural bimodality assumption made by the analytical model. The structural similarity of two model proteins is straightforwardly captured by the number of amino acid residue contacts that two structures have in common (Q-score, see *Methods*) (49).

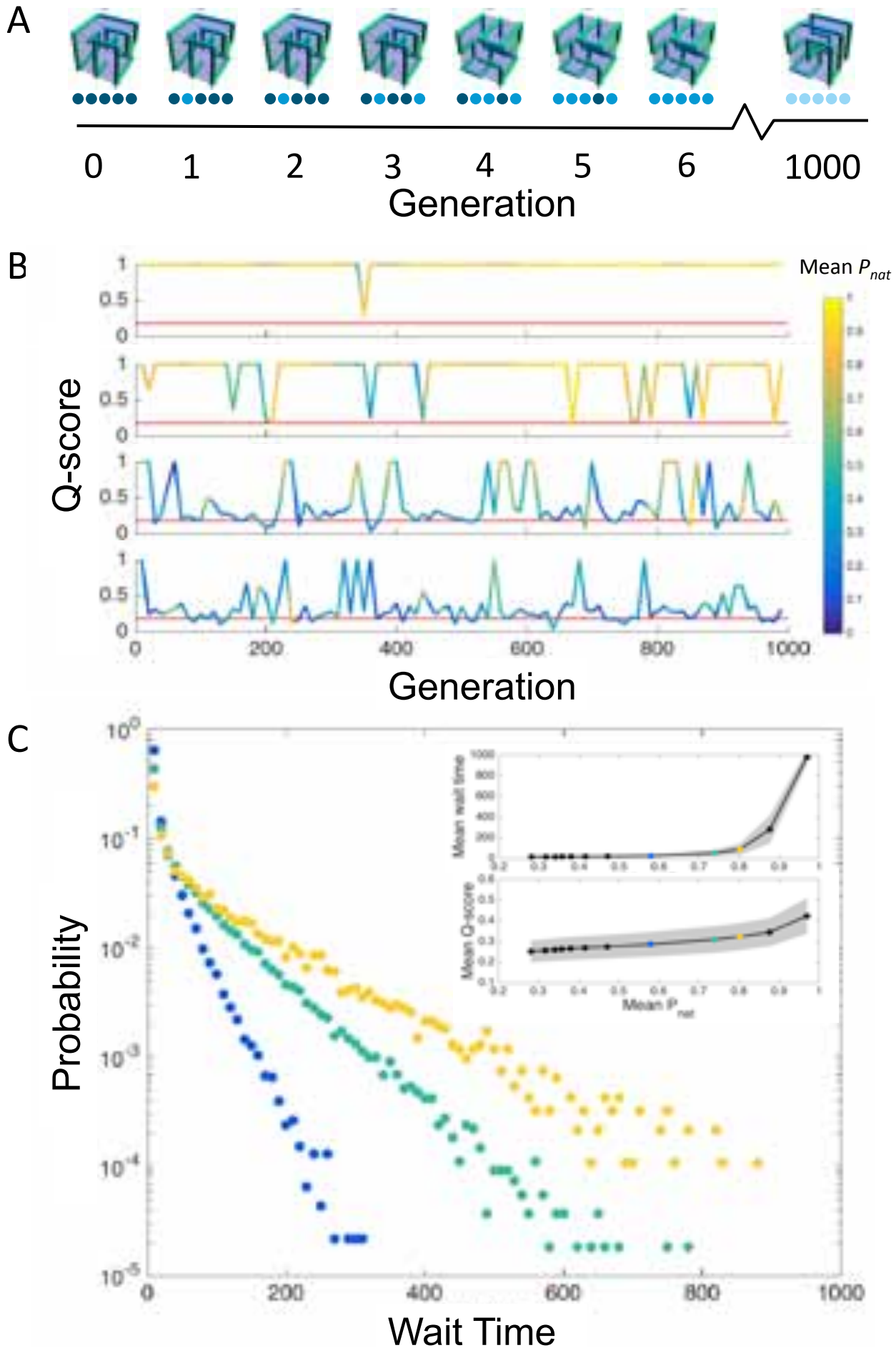

**Figure 3: Dynamics of fold evolution** (A) Schematic representation of Monte Carlo simulated evolution. Protein sequences (represented by a series of circles below each lattice structure, color indicates amino acid type) acquire mutations, periodically causing the protein to transition to a new structure. (B) Four representative fold evolution trajectories at different mean thermodynamic stabilities ($\bar{P}_{nat}$). Dips indicate structure evolution events. Trajectory color illustrates the thermodynamics stability of the evolving proteins, as described in the color bar. (C) The distribution of wait times between fold discovery events for three different $\bar{P}_{nat}$, $\bar{P}_{nat} = 0.58$ (blue), $\bar{P}_{nat} = 0.74$ (green), and $\bar{P}_{nat} = 0.80$ (yellow). In the insets and in (C), the colored data points also correspond to these $\bar{P}_{nat}$. The top inset shows the average wait time between fold discovery events for the different values of $\bar{P}_{nat}$ tested. The bottom inset shows the average structural similarity (Q-score) between the previous structure and the arising structure for fold discovery events.

In Figure 3B, representative individual trajectories from four selection regimes are plotted such that protein structure discovery events are reflected as dips in Q-score, and protein stability is indicated by color. The average Q-score, $\bar{Q}$, between a random pair of model proteins is $0.19 \pm 0.08$, as indicated by the red line in Fig. 3B. For each tested selection pressure, an average Q-scores associated with fold evolution events, $\bar{Q}_i$, can be determined from the simulation data by calculating the Q-score between the ancestral and mutant fold each time a new fold arises, and averaging over these values. As depicted in the bottom inset of Fig. 3C, $\bar{Q}_i$ does not differ significantly from $\bar{Q}$ when proteins evolve at the lowest $\bar{P}_{nat}$ observed ($\bar{P}_{nat} = 0.23$). The case is very different, however, when proteins evolve at their most stable $\bar{P}_{nat}$ observed ($\bar{P}_{nat} = 0.97$). Stable proteins are biased towards discovering new structures similar to ancestral structures such that $\bar{Q}_i = 0.42 \pm 0.17$, almost three standard deviations above $\bar{Q}$. The positive correlation between $\bar{Q}_i$ and $\bar{P}_{nat}$ likely reflects that discovering a mutant fold similar to the wild-type (high $Q$) avoids destabilizing evolutionary intermediates (Fig. S6). However, even strong selection pressure does not increase $\bar{Q}_i$ above ~0.42 because there exist so few structures with $Q > 0.42$ for any given evolving protein structure (Fig. S7) that their discovery appears unlikely for entropic reasons. The underlying rarity of similar structures in the space of model protein structures erects evolutionary barriers to fold evolution and limits the capacity of structure evolution to occur incrementally. This is reflected in the distribution $P(Q_{t,t+10})$, the probability distribution of Q-scores between protein structures at time $t$ and at time $t+10$, which is bimodal with peaks at $Q = 1$ and $Q \approx 0.29$ (See below, Fig. 4A-C), consistent with the bimodal distribution of TM-align scores discussed above.

A key assumption of the analytical model was that fold evolution is an activated process(50). We tested this assumption by examining whether wait times between fold discovery events were distributed exponentially, which is the hallmark of an activated process, and found that indeed, model protein evolution does follow activated dynamics (Fig. 3C).

Alleviating purifying selection for stability accelerates the rate of structure evolution by allowing proteins to maintain lower stability on average (lower $\bar{P}_{nat}$). The average wait time between structure discovery events diminishes rapidly as simulations are run at higher $T_{sel}$ such that the stability of the evolving proteins diminishes. When $\bar{P}_{nat} \leq 0.74$ selection pressure was so weak that nearly every recorded generation explored a different fold (Fig. 3B, Fig. 3C top insert), akin to the $k \approx 1$ regime in the analytical model. The abrupt transition from diffusive to activated dynamics is also apparent in the distribution of fold discovery events (Fig. S8A) which shows that the distribution is Poisson when $\bar{P}_{nat} \geq 0.74$, where structure evolution events are rare, and that the mean number of fold discovery events increases when $\bar{P}_{nat}$ falls below 0.74 (Fig. S8A, Dataset S2). In summary, structure evolution of model proteins follows activated kinetics in the strong selection regime and newly discovered structures appear much different from parent ones, providing strong supports of the postulates of the analytical theory.

Now we determine whether structure-sequence relationship reproduces the cusp shapes observed in real proteins (Fig.1). For unstable proteins, the pace of structure evolution outstrips sequence evolution leading to a concave sequence-structure relationship (Fig. 4A), which is consistent with the $k \approx 1$ regime of the analytical model and not observed in real proteins. The most interesting cases are at intermediate selection regimes where there is a sigmoidal cusp-like transition from high average structural similarity to low structural similarity at higher sequence divergence (Fig 4B-C). In this selection regime, proteins dwell in a particular structure while accumulating sequence mutations, yet are periodically, at longer time scales, able to transition to another structure, which is substantially different from the preceding one. For extremely stable proteins, by contrast, sequence can still evolve readily, but structure evolution is severely hindered (Fig. 4D). The increased dwell-time in a particular protein structure also allows sequences adopting that fold to proliferate over time, which is reflected in the correlation between $\overline{P}_{nat}$ and the mean structure family size (Figure 4E). Structure family size is equivalent to gene family size, discussed in Shakhnovich et al., except that amino acid sequences rather than nucleotide sequences are considered here (41).

The analytical model and simulations indicate that a divergent scenario reproduces the key features of the sequence-structure relationships shown in Fig.1. In this model, the plateau at high structural similarity arises from evolutionary fitness barriers impeding structure evolution, not from a biophysical fact that sequences encoding stable proteins and sharing more than 30% of their residues must encode the same structure. So far, however, we have not been able to fully exclude the latter possibility. In order to test whether a purely biophysical constraint might explain the cusped sequence-structure relationship, we simulate an alternative, convergent, mechanism of structure discovery. To simulate convergent evolution, proteins start from sequences that stably fold into a randomly chosen structures but the fitness function favors sequence convergence to another, "target" protein (see *Methods* for details). This scheme was constructed to test the pure biophysics hypothesis of the cusp's origin, described above, and obviously does not reflect actual convergent evolutionary forces found in nature. In Figures 4A and 4C we accompany the divergent evolution results described above with the results of convergent evolution simulations, shown in gray. Under weak selection for folding stability, the convergence of protein structure as sequences evolve follows a similar path as it does during divergent evolution (Fig. 4A). By contrast, under strong selection for stability, protein structures begin to converge at a much higher sequence identity than where they begin to diverge (approximately 85% and 25% respectively, Fig. 4C). Proteins converging under the constraint of stability also attain only $69 \pm 9\%$ sequence identity with respect to the target sequence after 3,000 mutation attempts and only 0.5% of the evolving proteins attained the target structure, compared to 100% in the weak selection regime. Therefore, evolutionary dynamics must play a role in generating the sequence-structure cusp and a pure biophysics explanation is rejected.

Taken together, our results from divergent and convergent evolution scenarios indicate that selection pressure generates a peculiar hysteresis in evolutionary trajectories: When proteins diverge under selection for folding stability, they long retain the same structure because evolving a new structure involves passing through a fitness valley. Were two highly diverged (different sequence and structure) proteins to converge on the same sequence again, they would long retain *different* structures for the same reason. Such a scheme was realized experimentally in (18) in which two proteins with different sequences and structures were subjected to engineered single point mutations which increased their sequence identity up to 88% yet did not unfold the proteins or change their original structures.

**Population size modulating selection pressure**

Evolutionary simulations of individual model proteins, while biophysically realistic, lack biological realism: the concepts of fitness and selection are applicable to whole cells and populations rather than individual proteins. To address this shortcoming, we performed simulations of an evolving population of competing single-cell model organisms, as depicted in Figure 5A and described in detail in *Methods* and (51). Model organisms have genes that encode 27-mer lattice model proteins. Each generation, an organism can die, divide, undergo a gene duplication event, or undergo a genetic point mutation. This evolutionary scheme mimics the natural emergence of protein families and was previously used to explain the power-law distribution of protein family and superfamily sizes observed in nature (51). In contrast to the previous model, selection pressure is here applied to the whole organisms rather than only the evolving proteins, which makes the situation more biologically realistic and non-trivial. The strength of selection (proxied by $T_{sel}$ in the previous model of individual protein evolution) is determined by the population size (52, 53) in the present, more biologically realistic model.

Evolutionary runs were simulated over a range of maximum population sizes, $N_{\max}$, from 500 to 5,000 organisms, in replicates of 50. Whenever birth of new organisms drives the population size, $N$, above $N_{\max}$, $N-N_{\max}$ randomly selected organisms are removed to ensure constant population size, simulating a turbidostat (51).

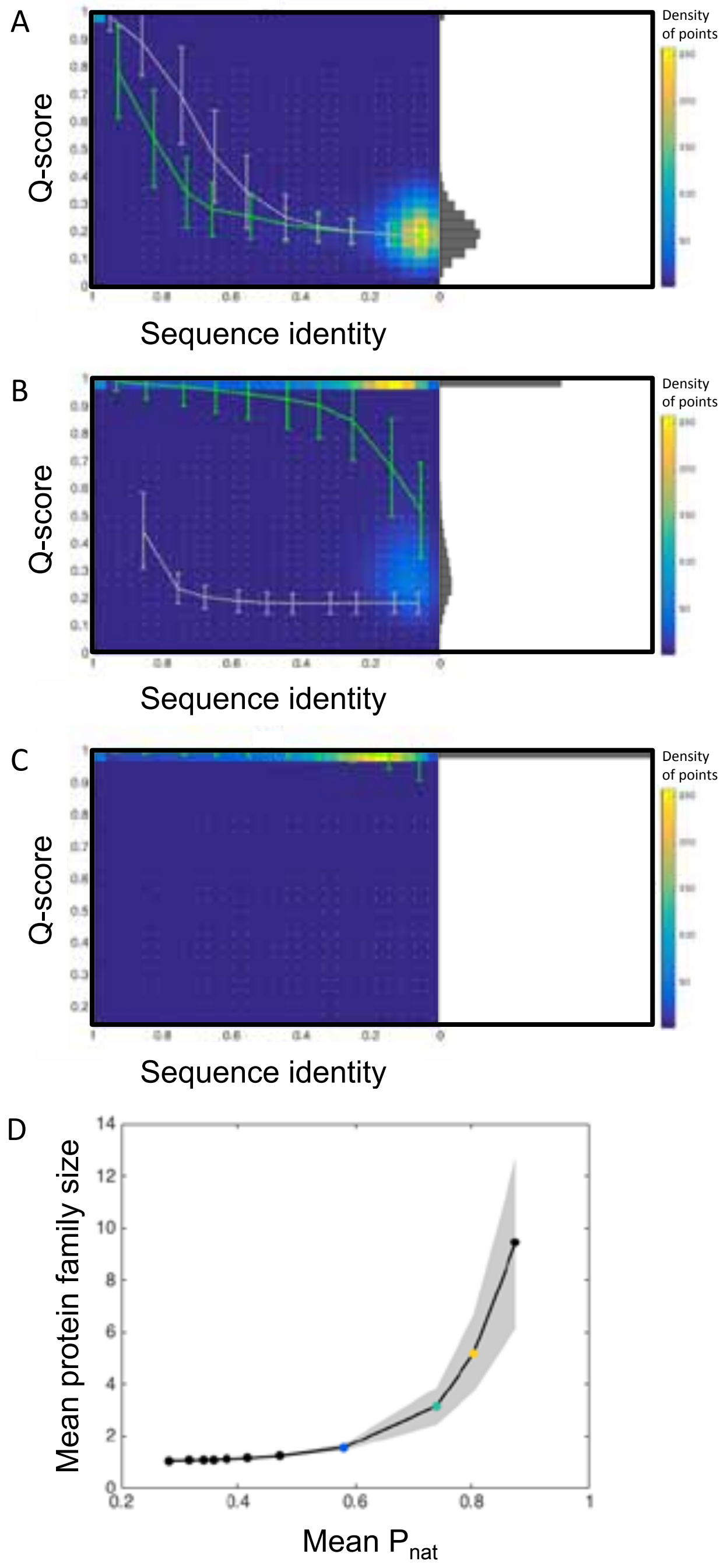

**Figure 4: The relationship between sequence divergence and structure evolution for model proteins** (A-C) The sequence identity and Q-score for each pair of proteins compared. White points indicate the data points from proteins arising in the same trajectory. The color scale indicates the density of points for each pair of Q-score, sequence identity values, and the average Q-score in each sequence identity bin of size 0.1 is calculated and plotted in green for divergent evolution simulations and in grey for convergent evolution simulations (see Methods) The histogram represents the frequency of Q-score pairs. (A) Weak selection for folding stability: $\overline{P}_{nat} = 0.32$. In this regime structure diverges more rapidly than sequence. The result of convergent evolution simulations without selection for stability is shown in grey. There is not significant evolutionary hysteresis when selection for folding stability is low as can be seen in the similarity of the green and grey curves. (B) $\overline{P}_{nat} = 0.88$ When proteins evolve under strong selection for folding stability, as they do in both the divergent (green) and the convergent (grey) evolution simulations, there is clear evolutionary hysteresis: depending on whether a pair of proteins with 50% sequence identity are diverging or converging their structure will most likely be extremely similar or very different respectively). (C) $\overline{P}_{nat} = 0.97$ : At such strong selection for thermodynamic stability, structure evolution is limited because stability valleys separating structures are not overcome on the time scale of simulations. (D) The mean structure family size as a function of $\overline{P}_{nat}$. The size of a structure family is the number of non-redundant sequences (sequence identity < 0.25) evolved in simulations that adopt a particular structure. Shaded regions show one standard deviation and colored points mark $\overline{P}_{nat} = 0.58$ (blue), $\overline{P}_{nat} = 0.74$ (green), and $\overline{P}_{nat} = 0.80$ (yellow).

After many generations, we recorded the extant proteins from all organisms in a particular evolutionary run, calculated their stabilities, and calculated the sequence identity and structural similarity (Q-score) of all pairs of proteins. We found that the evolutionary runs yielding unstable proteins have qualitatively different relationships between sequence and structure divergence than other replicas. In these cases, the extremely rapid turn-over of structures is manifested in a concave-up dependence of average Q-score on sequence identity (green and blue trajectories in Fig. 5B), as in the low $P_{nat}$ regime of the Monte Carlo simulations (Fig. 4A). Once the population size passes the critical threshold, of about 1,250 organisms, the characteristically cusped sequence-structure relationships become the most probable result of evolution (Fig. 5B, middle and bottom). On the other hand, we do not see a population size where structure evolution is almost entirely shut down, even at the largest population sizes probed in simulations. This could reflect that in larger populations, there is a larger influx of beneficial as well as deleterious mutations, so more neutral or beneficial fold evolution events can occur.

We found that the average $P_{nat}$ of proteins at the end of 3,000 generations ($\bar{P}_{nat}$) strongly anti-correlates with the diversity of structures observed in the population, as reflected in the structure entropy:

$$H = -\sum_{i=1}^{10,000} p_i \log(p_i) \quad (4)$$

where $p_i$ is the probability of finding structure $i$ in the final generation of the simulations (inset of Fig. 5C). The correlation between $\bar{P}_{nat}$ and $H$ reflects the fact that less stable proteins have faster rates of structure evolution. This correlation does not depend on population size (different colors in Fig 5B inset). Rather, we found that population size modulated the number of evolutionary replicas that failed to evolve stable proteins (Fig. 5C). This is also reflected in the relative peak heights of the Q-scores shown in Figure 5B.

Finally we examine the size of the average protein structure family across the range of $\bar{P}_{nat}$ explored in the simulations (Figure 5D). The size of protein structure family is defined as the number of genes encoding non-redundant protein sequences (sequence identity < 25%) but whose gene products adopt the same native state structure. In datasets of natural proteins, protein structure families of various sizes have been observed and overall, there is a positive correlation between gene family size and protein contact density (41). In the context of these simulations, which instantiate the crucial mechanisms of structure family creation and growth (sequence evolution, gene duplication, and structure evolution) we observe a strong positive correlation between $\bar{P}_{nat}$ of proteins at the end of an evolutionary replica and the average structure family size (Fig. 5D). The significance of the trend is further magnified by the observation that the total number of genes in a population at the end of a simulation actually correlates negatively with $\bar{P}_{nat}$ (Fig. S9). Overall, this observation supports the view that stable proteins are trapped in particular structures, providing more time for the number of sequences adopting this structure to grow.

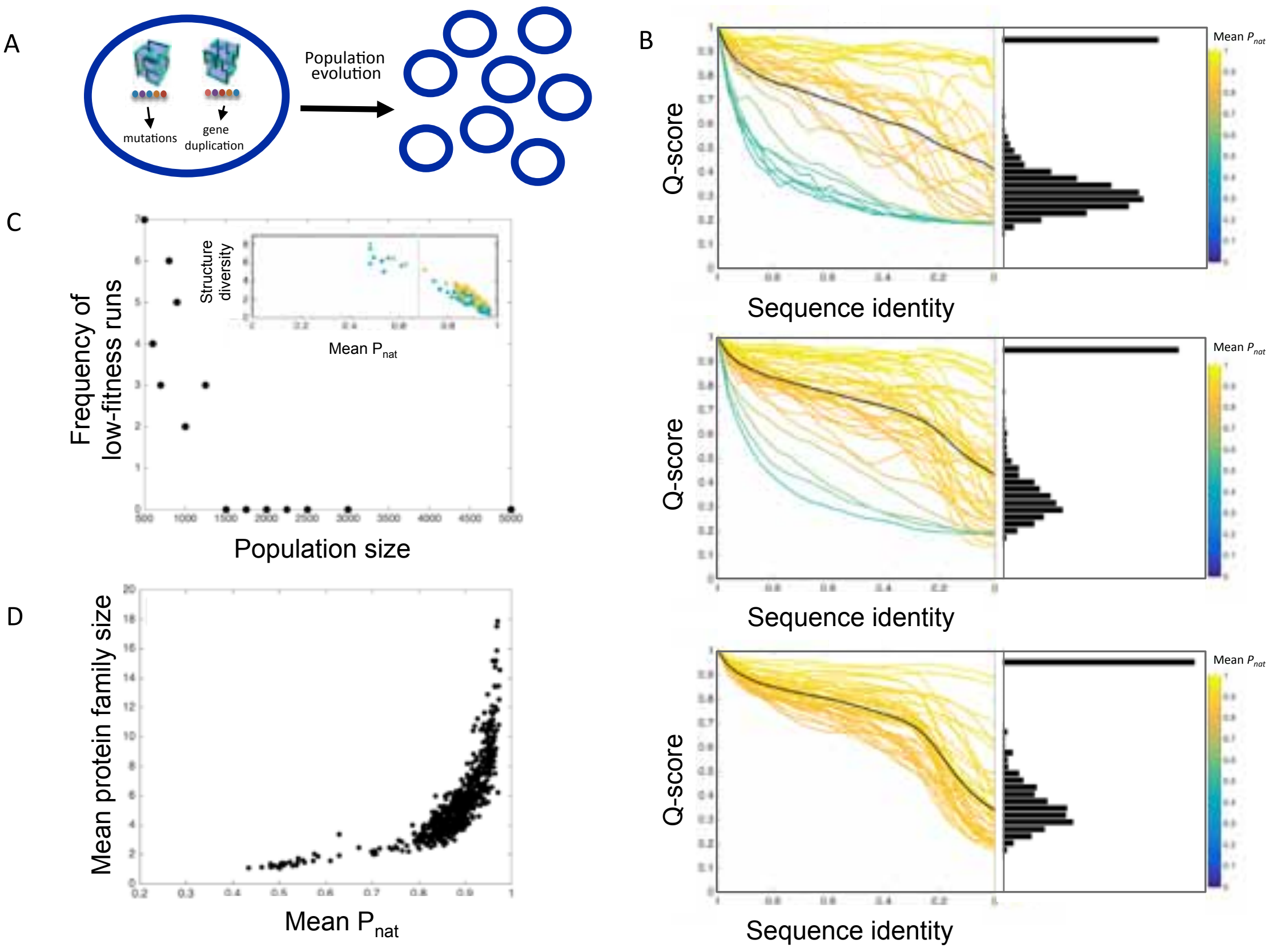

**Figure 5: Population size modulates selection pressure** (A) Schematic representation of the multi-scale evolution model. Each organism consists of genes that code for proteins and can be duplicated or acquire point mutations. Organisms evolve under selection pressure for protein folding stability. At the end of the simulation, sequence identity and Q-score are calculated for each pair of extant proteins in the population. (B) The number of replicas that evolved the full 3,000 generations yet failed to evolve stable proteins ( $\overline{P}_{nat} < 0.68$ ) for each population size. Notably, above population sizes of 1,250 the selection pressure is apparently strong enough to drive evolution of stable proteins in each replica. Inset: the mean $P_{nat}$ of proteins at the end of an evolutionary trajectory correlates with the diversity of structures. Dark green: population size =500, light green: population size = 1,250 and yellow: population size = 5,000 (C) The relationship between sequence divergence and structure evolution in population simulations. For each trajectory, the average Q-scores are plotted in a color indicating $\overline{P}_{nat}$ of proteins at the end of the simulation. The black line is the average over the individual trajectories. Top panel: Population size = 500. Middle panel: Population size = 1,250, Bottom panel: Population size = 5,000. Histograms showing the bimodal distribution of Q-scores are shown for each population size. Consistent with the results described above, in the realistic context of population dynamics, the characteristic cusp-like dependence of structural similarity (Q-score) on sequence identity only emerges at strong selection pressure. (D) Mean structure family size as a function of $\overline{P}_{nat}$ of proteins at the end of each trajectory.

**Figure 5 cont'd: Population size modulates selection pressure** Furthermore, strong evolutionary pressure increases relative height of the high Q-score peak in the bimodal Q-score distribution relative to the low Q-score peak. (D) Strong evolutionary pressure increases the number of generations between fold discovery events and because it generates more stable proteins, is also likely to lead to larger structure family sizes through gene duplication in this model. Indeed, the largest protein families arise in simulations where evolved model proteins have a high mean $P_{nat}$ (r=0.9, p=0).

## Discussion

We presented a simple physical analytical model of protein structure evolution that explains why there is a cusped relationship between structure and sequence divergence. Under the constraint of protein folding stability, fitness valleys form barriers that separate sequences encoding stable protein structures. The most stable proteins face formidable fitness valley barriers and therefore, slow structure evolution rates. Continuous sequence evolution degrades sequence identity of diverging proteins over the timescale needed to accumulate the mutations that traverse these valleys. Our simulation results show that protein stabilities and their accompanying rates of structure evolution, $k$, arise as a result of differential selection pressures for stability. Strong selection for stability causes proteins to evolve high stability and hinders structure evolution, while weak selection permits rapid exploration of structure space, of predominantly unstable proteins. Interestingly, this observation strengthens the analogy to chemical kinetics: just as the ratio of the free energy barrier to temperature controls the rate of barrier crossing in chemical kinetics, the ratio of the fitness barrier to the appropriate measure of strength of evolutionary selection controls the rate of structure discovery in the evolutionary context of the model (52).

Importantly, our bioinformatics analysis of protein domains in SCOP shows that the effect of selection strength, as reflected in proteins contact densities, is not just theoretical but actually modulates the relationship between structure and sequence divergence, which was previously thought to be universal. The analytical model we propose not only explains the existence of the cusp, it also recapitulates and explains why the transition to the twilight zone becomes sharper as selection pressure increases: high selection decreases the probability that a new fold will emerge before decay of sequence similarity saturates.

Our bioinformatics results were the same when we tested a measure of sequence similarity that accounts for similarity of physical properties between substituted amino acids. While no other study has examined the effect of contact density on the sequence-structure relationship, Wilson et al. rigorously tested multiple methods of scoring sequence similarity (including percent sequence identity, Smith-Waterman Score, and statistical significance of sequence similarity) and confirmed that the non-linear dependence of structure divergence on sequence divergence is independent of the methods used (5). Interestingly, several studies that focused on the structure evolution that occurs only within protein families, thereby excluding the distantly related proteins included in Wilson 2000 and in our study, reported that the non-linearity does not arise consistently when percent sequence identity was substituted with other, more advanced, measures of sequence similarity (7, 8, 54). This apparent contradiction likely arises from the close evolutionary relationship among proteins in a protein family, which share a common ancestry as inferred by significant sequence similarity, very similar structure and function. Sequence identity saturates more rapidly than other measures of sequence similarity so when only closely related proteins are examined, sequence identity begins to saturate, causing apparent non-linearity, while others such as NW-score or bitscore may or may not begin to saturate depending on the family (8). Therefore, using a measure of sequence similarity such as statistical significance of the similarity generally yields a linear relationship *within protein families* (7) However when all proteins sharing a particular SCOP fold are examined (5) the relationship between sequence and structure divergence remains non-linear, with the cusp present irrespective of the particular definition of sequence and structure divergence (z-scores and P-values were used as measures of statistical significance in (7) and (5) respectively ). We also note that this finding is consistent with our divergent evolution model, where non-linearity emerges from rare evolutionary transitions between significantly different protein structures (Fig. 4). Thus by limiting the analysis to protein families (7) and (8) excluded significantly diverged structures and, not surprisingly, observed linear relationships between structure and sequence divergence for most families. By contrast, we were able to analyze and explain the non-linearity in the sequence-structure relationship by expanding the analysis to the level of protein fold.

An underlying motivation of these previous works was to understand how a protein's structure is encoded in its residues, namely, whether structural information is encoded globally across all residues or whether it is localized to a subset of "gatekeeper" residues (55-58). Wood and Pearson argued that a non-linear relationship between sequence and structure divergence was only consistent with the latter mechanism (7, 8). Indeed, there is a very clear mechanism by which the gatekeeper model would generate a cusped relationship between sequence and structure divergence: if, for example, only 30% of residues determine protein structure, then 70% can evolve without disrupting protein structure, and it is not until these gatekeeper residues accumulate mutations that a new structure emerges. Our analytical model, however, is a global model of residues determining protein structure because we do not define any privileged gatekeeper residues. Thus, we clearly demonstrate that a global model based on fitness barrier crossing is also consistent with the cusp and twilight zone.

In order to test whether a local model whereby few gatekeeper residues determine protein structure might also consistent with the bioinformatics and simulation data, we constructed a second analytical model based on this mechanism (see SI for derivation). This model predicts that the position of the twilight zone depends on the number of gatekeeper residues (Fig. S10). While this result may be intuitive, it is not consistent with the bioinformatics or simulation data, even though the 30% position of the twilight zone is not imposed on these in any way. That we do not observe a moving twilight zone in the data indicates either that folds have roughly the same number of gatekeeper residues regardless of stability or that structure evolution is mediated globally rather than via a few key residues. Contrary to previous work, therefore, we have shown that the global model is not only consistent with the bioinformatics data but is actually better at explaining the bioinformatics data than the previously favored local model.

Despite its simplicity, the proposed structure evolution mechanism is powerful in placing many longstanding observations in protein biophysical evolution in a single interpretive framework. Along with tuning the rate of structure evolution, another effect of selection is to modulate the size of sequence families, i.e. families of sequences that fold into the same native structure. This is because strong selection slows the pace of structure evolution more dramatically than sequence evolution, so during the time period a protein is "trapped" in a particular fold by selection, continued duplication and exploration of the sequences adopting that fold generates larger and larger sequence families.

It has been clear for years that protein contact density is deeply connected to evolution. Proteins with high contact density tend to be old, part of larger gene families, and part of larger structural neighborhoods (defined as the number of non-redundant sequences adopting similar, but different structures, analogous to a SCOP fold) (41, 59). The root of these observation has been attributed to intrinsic evolvability conferred upon a protein by its contact density, and it has been hypothesized that young proteins may evolve quickly because they are under positive selection to evolve stability of new functions (41, 59). Our work provides an alternative framework within which to interpret these findings. We suggest that pressure for folding stability that causes proteins to evolve high contact density, severely constraints structure evolution, leads to larger gene families due to longer divergence times between structure innovation events, and selects for discovery of structures similar to the ancestral structure, to the extent that its possible, which leads to larger structural neighborhoods.

It may seem counterintuitive that stability retards the rate of structure evolution because in the context of directed evolution, stability typically enhances evolvability (60). Stability promotes evolvability during directed evolution because engineered stabilizing mutations create a stability buffer that allows the protein to tolerate destabilizing mutations that confer a new function *without changing the structure*. However in the context of natural evolution, a protein's stability reflects mutation-selection balance, the point at which selection for protein folding stability is balanced by mutational pressure towards less stable sequences (30). Therefore proteins that are naturally very stable are such because they are under stronger pressure for stability (e.g. they may be more abundant in the cytoplasm) and for that reason they might not have a reservoir of stability that can be used up and regenerated during for subsequent rounds of structure evolution (61). Supporting this point, we observed that structure evolution events were associated with loss of stability for proteins of marginal stability, but not for the most stable proteins (Fig. S11).

Here, we reported a negative correlation between contact density and structure evolution rate. Curiously, when focusing on contact density and *sequence* evolutionary rate, Zhou et al. reported a positive correlation (62). The influence of selection on contact density and evolutionary rates therefore apparently depends on the type of evolutionary rate examined (structure versus sequence) and on the timescales (long versus short). Overall, we view contact density and its associated metrics as neither a sign of intrinsic evolvability nor evolution under weak selection, as most previous studies have, but rather as a signature of strong evolutionary selection (41, 59, 62). These interpretations may not be diametrically opposed but further studies may be needed to clarify their relationship.

## Methods

### Structure divergence of proteins sharing a common SCOP fold

We use the set of all α, β, α+β, and α/β protein domains in SCOP including mutants (25). The contact density of each domain was calculated as a predictor of thermodynamic stability. When calculating contact density, we consider two residues in contact if any of their non-hydrogen residues are within 4.5 angstroms of each other. Proteins belonging to top (>4.93 contacts/residue), middle (4.57<CD<4.65 contacts/residue), and bottom 10% (<4.13 contacts/residue) with respect to contact density were studied further.

Following Chothia and Lesk (4), who studied the divergence of proteins within protein families, we studied the divergence of proteins classified into the same SCOP fold. By choosing a broader classification than family, we could track divergence over the long timescales over which significant structure evolution takes place. SCOP folds are extremely broadly defined so proteins sharing a fold classification often adopt significantly different structures. The structural similarity of protein domains classified in the same SCOP fold, in the same contact density class (both having high, intermediate, or low contact density) and comparable length (within 10 amino acids) were compared using the Template Modeling-align (TM-align) algorithm which yields a number representative of the similarity between two folds, ranging from zero to one (63). Then their percent sequence identity was calculated from their alignment by the Needleman-Wunsch algorithm and Blosum30 substitution matrix, implemented in MATLAB. Proteins classified in different SCOP folds do not share significant sequence homology or structural similarity and were therefore not compared. In Figures 1B-D and the corresponding Dataset 1, the data are subdivided by contact density alone, not by protein class or fold.

**Fitting analytical model to bioinformatics data**

In order to quantify the differences in structure evolutionary rates of real proteins apparent in Figure 1, we fit the analytical expression for structure evolution, equation **Error! Reference source not found.**, to the data presented in Figure 1B-D (where sequence and structure similarity is subdivided by contact density) as well as in Figure S1 (where the data is subdivided by structural class). We proceeded by binning the data into 50 bins spanning sequence identity of zero to one. This step was necessary to achieve a fit to the data because otherwise the large majority of data points at low sequence identity and low structural similarity dominate and foreclose the possibility of a meaningful fit to the high sequence identity and cusp regions of the data. Because the twilight zone of the analytical model occurs at 5% sequence identity, the average sequence identity between two random sequences, while the bulk of proteins compared in the twilight zone share 20% sequence identity for real proteins, possibly reflecting that sequence alignment algorithms seek to maximize the overlap of sequence pairs being aligned, we added a parameter $c$ to the equation in order to shift the analytical model twilight zone to the twilight zone of the data. The exact curve that was fitted to the data was

$$\tilde{S}(q)=1-\frac{1-a}{a}\left(\left(1-\frac{a}{l(a-1)}\right)^{\frac{1-q}{k}}-1\right)+c \qquad (5)$$

where $\tilde{S}(q)$ is the sequence identity as a function of structure survival probability, a rearrangement of Eq. 3 plus $c$, the twilight zone shift. The parameter $a$ is 20, the number of amino acid types, and $l$ is the protein length, which is set to the average length of proteins in the dataset being fit. $\tilde{S}(q)$ rather than $q(S)$ was fit to the data in order to account for the twilight zone in more detail when using the binning procedure. Eq. 5 was fit to the binned bioinformatics data using the program Igor, which optimized the parameters $k$ and $c$ for fit.

**Protein model**

We use standard 27-mer lattice-model proteins to simulate the structural evolution of proteins (64). Lattice proteins can fold into 103,346 fully compact structures for the 3x3x3 cubic lattice, and following (45), we use a representative subset of 10,000 randomly chosen structures for computational efficiency. All proteins in this model have 28 contacts and therefore have identical contact densities. There are 20 amino acid types in the model. The energy of a protein in any given structure can be computed from the Miyazawa Jernigan (MJ) potential (46), which contains interaction energies for each spatially proximal pair of any two amino acid types. The energy of a 27-mer sequence adopting a particular structure, $S$, is given by

$$E_S = \frac{1}{2}\sum_{ij=1}^{27} C_{ij}^{(S)} \mathrm{M}_{AA(i)AA(j)} \tag{6}$$

Where $C^{(S)}$ is the contact matrix of structure $S$, whose elements $C_{ij}^{(S)} = 1$ when residues $i$ and $j$ form a non-covalent contact and $C_{ij}^{(S)} = 0$ otherwise. $M$ is the MJ interaction energy matrix such that the element $M_{AA(i)AA(j)}$ contains the interaction energy of the amino acid types of residues $i$ and $j$. The Boltzmann probability, $P_{nat}(T)$, of a protein adopting its native state (lowest energy) structure is therefore given by the canonical partition function:

$$P_{nat}(T) = \frac{e^{-E_{nat}/T}}{\sum_{i=1}^{10,000} e^{-E_i/T}} \tag{7}$$

Where $E_i$ is the protein's energy in structure $i$ and $T$ is the temperature in arbitrary units. The structure in which a protein's energy is minimized is considered the protein's native state.

The degree of structural homology between two model proteins is quantified using the number of contacts the structures have in common, normalized by the total number of contacts (28 contacts for all compact 27-mer model proteins) (49, 63).

$$Q_{1,2} = \frac{1}{28}\sum_{ij=1}^{27} C_{ij}^{(1)} C_{ij}^{(2)} \tag{8}$$

Where $C^{(1)}$ and $C^{(2)}$ are the contact matrices of the two structures being compared such that $C_{ij} = 1$ if residues $i$ and $j$ are in contact and 0 otherwise, excluding covalently linked residues as in Eq. 6.

**Monte Carlo algorithm for divergent evolution**

The model proteins were evolved under selection for folding stability, $P_{nat}$. This constraint captures two biological features. First, most proteins must be folded to carry out their function (the obvious exception is intrinsically disordered proteins). Second, unfolded or misfolded species can be toxic (40, 65, 66). Each simulation of model protein evolution proceeded in two steps. First, each simulation was initialized with a protein stably adopting a particular structure ($P_{nat} > 0.99$). Stable proteins were made by generating a random 27-mer amino acid sequences, introducing random mutations, and accepting the mutations only if they stabilized a predetermined ground state (67). This procedure, rather than a Monte Carlo procedure, is sufficient for generating unique, stable sequences for each structure. Then, evolutionary trajectories at different selection pressures for folding stability were carried out as follows. Each generation, the protein was subjected to a point mutation. The fitness effect of the point mutation was defined as follows:

$$f = P_{nat}^{(trial)} - P_{nat}^{(original)} \tag{9}$$

Where $P_{nat}^{(trial)}$ is the stability of the protein with the mutation and $P_{nat}^{(original)}$ is the stability of the protein prior to the mutation. Any neutral mutation or mutation increasing fitness was accepted while destabilizing mutations were accepted or rejected according to the Metropolis criterion with $e^{f/T}$. The selective temperature, controls the stringency of selection for protein stability; the more destabilizing a mutation, the less likely it is to be accepted (67). $T_{sel}$ directly tunes the equilibrium $P_{nat}$ of evolving proteins (SI, Fig. 1 and Table 1). Simulations ran for 1,000 mutation attempts, illustrated in Fig. 3A and the structure, sequence, and stability ($P_{nat}$) were recorded every 10 generations. This was repeated for approximately 2,750 different starting structures at different strengths of selection for stability $P_{nat}$ (set by $T_{sel}$).

**Monte Carlo algorithm for convergent evolution**

In these simulations, many proteins all initially adopting different structures with high $\overline{P}_{nat}$ evolve under selection pressure to maximize their sequence identities with respect to the same designated "target" sequence. The target sequence is a randomly selected sequence with a thermodynamically stable ground state. To simulate the regime where selection pressure for folding stability is low (Figure 4A), any mutation increasing sequence identity with respect to the target sequence ($SID_{i,\text{target}}$) was accepted. No selection pressure for folding stability was applied. The regime where selection pressure for folding stability is high was simulated by using the fitness function $SID_{i,\text{target}} \times P_{nat}$ (Figure 4C). A single convergent evolution simulation was run in each selection regime, consisting of 100 proteins converging over 3,000 generations in the weak selection regime while 200 protein converging over 3,000 generation were used in the strong selection regime to compensate for slower evolution and improve statistics. To determine the sequence-structure relationship for convergent evolution, the sequence identity and Q-score of the converging proteins were calculated every 100 generations within each regime.

**Multi-scale modeling of protein evolution**

We used the multi-scale simulation algorithm developed and described in detail in Zeldovich et al. 2007 (51). Because this evolutionary procedure yields protein families of sizes matching the distributions observed in nature, it is an ideal source of comparison with SCOP data. The simulations were initialized with 100 identical organisms, each consisting of a single gene (81-mer DNA) with random sequence and its 27-mer gene product. (Thus, an average initial protein has a $P_{nat}^{rand} = 0.23$ corresponding to average stability of random sequences) At each step in evolutionary time, one of five fates affects each organism with equal probability (1) no event (2) point mutation at a random position in the genome to a random nucleotide at rate $m = 0.3$ per unit time per base pair (3) gene duplication with rate $\mu = 0.03$ of a random gene in the organism's genome (4) death of the organism with rate $d$ as given by $d = d_0(1 - \min_i P_{nat}^{(i)})$, and (5) division of the organism into two equivalent daughter cell with birthrate $b = 0.15$. Thus, for example, there is a 20% chance that gene duplication is selected as an organism's possible fate and a further probability $u$ that a gene will actually be duplicated. As the population evolves, the population size may increase until it reaches a carrying capacity, $N_{\max}$. Once the maximum population size is attained it is maintained by removing $N - N_{\max}$ randomly chosen organisms from the population each generation. We modulate the strength of selection in the

population by simulating populations with size $N_{\max}$ set at 500, 600, 700, 800, 900, 1000, 1250, 1500, 1750, 2000, 2250, 2500, 3000, and 5000.

Fifty evolutionary trajectories were run at each population size. As the simulation progresses, proteins evolve, becoming more stable and periodically changing structure, spawning new protein families. Trajectories were terminated after 3,000 generations or when the population went extinct. Only simulations that ran the full 3,000 generations were analyzed, and even among these evolutionary runs, in some cases, the proteins did not evolve high $\overline{P}_{nat}$ (Fig. 5B-C). For our analysis of structure divergence, we calculate the pairwise sequence identity and Q-score of each extant pair of proteins harvested from a particular evolutionary run.

**Author contributions**

A.I.G and E.I.S. designed the research, A.I.G. carried out the bioinformatics analysis and developed and performed Monte Carlo simulations. A.M.C. performed the multi-scale evolution simulations and developed the analytical model under the guidance of A.I.G. Data was contributed and analyzed by J.M.C. A.I.G. and E.I.S. prepared the manuscript.

**Acknowledgements**

We thank the Kavli Institute for Theoretical Physics Quantitative Biology Summer School, which A.I.G. attended in 2015 and where she had many useful discussion, especially with Ned Wingreen and other participants in the discussion group organized by Tal Einav. This work was supported by NIH grant GM068670 and a National Science Foundation Graduate Research Fellowship (awarded to A.I.G.).

**Supporting Information**

# The role of evolutionary selection in the dynamics of protein structure evolution

Amy I. Gilson[1], Ahmee Marshall-Christensen[1], Jeong-Mo Choi[1] & Eugene I. Shakhnovich[1]

[1]Department of Chemistry and Chemical Biology, Harvard University, 12 Oxford St., Cambridge MA, USA

*Correspondence should be addressed to E.S. (shakhnovich@chemistry.harvard.edu)

**Derivation of the analytical model of protein evolution with structure evolution barriers**

We begin by deriving an expression for the Hamming distance, $S(t)$, between a protein at time $t$ and its ancestral sequence at time $t=0$, i.e. the number of derived rather than ancestral amino acid types comprising the protein at time $t$. A relationship for the expected Hamming distance between the evolving protein at time $t+1/\mu$ with respect to the ancestral protein is given by

$$S(t+1/\mu)=\frac{l-S(t)}{l}(S(t)+1)+\frac{S(t)}{l}\left(\frac{1}{a-1}(S(t)-1)+\frac{a-2}{a-1}S(t)\right) \quad \text{(S1)}$$

where $l$ is the length of the protein and $a$ is the number of amino acid types (typically 20) and $\mu$ is the mutation rate of the protein. Each of the three terms in Eq. S1 captures one of the three possible effects of a mutation. First, a mutation may increase $S(t)$ by one if it substitutes a residue in its ancestral state with a different amino acid type. Second, a mutation may revert a residue from a derived amino acid type back to its ancestral type, decreasing $S(t)$ by one. Finally, $S(t)$ remains constant if a mutation substitutes a residue with a derived amino acid type with another non-ancestral amino acid type. This expression simplifies to

$$S(t+1/\mu)=1+\frac{(a(l-1)-1)S(t)}{al} \quad \text{(S2)}$$

Setting $a_k=S(k/\mu)$, we obtained a mathematic sequence, a closed form for which can be found by using generating functions. Define $A(x):=\sum_{k=0}^{\infty}a_k x^k$. Noting that $a_0=S(0)=0$, we have

$$A(x)=\sum_{k=1}^{\infty}a_k x^k=\sum_{k=1}^{\infty}\left(1+\frac{(a(l-1)-1)a_{k-1}}{al}\right)x^k \quad \text{(S3)}$$

$$A(x)=\sum_{k=1}^{\infty}x^k+\sum_{k=1}^{\infty}\left(\frac{(a(l-1)-1)a_{k-1}}{al}\right)x^k \quad \text{(S4)}$$

$$A(x)=\sum_{k=1}^{\infty}x^k+\sum_{k=1}^{\infty}\left(\frac{(a(l-1)-1)}{al}\right)x\sum_{k=1}^{\infty}a_{k-1}x^{k-1} \quad \text{(S5)}$$

$$A(x)=\frac{x}{1-x}+\frac{a(l-1)-1}{al}xA(x) \quad \text{(S6)}$$

This implied the closed form $A(x)=\left(\dfrac{x}{1-x}\right)/\left(1-\dfrac{a(l-1)-1}{al}x\right)$ .

Expanding the function into a power series yields

$$A(x)=\sum_{k=0}^{\infty}\frac{al}{1+a}\left(1-\left(1-\frac{a+1}{al}\right)^{k}\right)x^{k} \tag{S7}$$

This implied $a_k=\dfrac{al}{1+a}\left(1-\left(1-\dfrac{a+1}{al}\right)^{k}\right)$, and recalling that $S(k/\mu)=S(t)$ , we concluded

$$S(t)=\frac{al}{1+a}\left(1-\left(1-\frac{a+1}{al}\right)^{\mu t}\right) \tag{S8}$$

Rearranging and setting $\mu=1$,

$$S(t)=\frac{l(a-1)}{a}\left(1-\left(1-\frac{a}{l(a-1)}\right)^{t}\right) \tag{S9}$$

We denote $k$ as the rate of structure evolution. Assuming exponentially distributed wait times between protein structure evolution events, the probability, $q(t)$, that an ancestral structure (the structure at time $t=0$) is unchanged at time $t$ follows immediately:

$$q(t)=e^{-kt} \tag{S10}$$

The expression for $S(t)$ can be rearranged to solve for $t$ which will permit us to write a parametric equation for structure evolution versus sequence evolution.

$$t=\frac{\log\left(1-S(t)\dfrac{a}{l(a-1)}\right)}{\log\left(1-\dfrac{a}{l(a-1)}\right)} \tag{S11}$$

Thus, the expression for structure evolution with respect to sequence divergence is

$$q(S) = \exp\left(-k\frac{\log\left(1 - S\frac{a}{l(a-1)}\right)}{\log\left(1 - \frac{a}{l(a-1)}\right)}\right) \tag{S12}$$

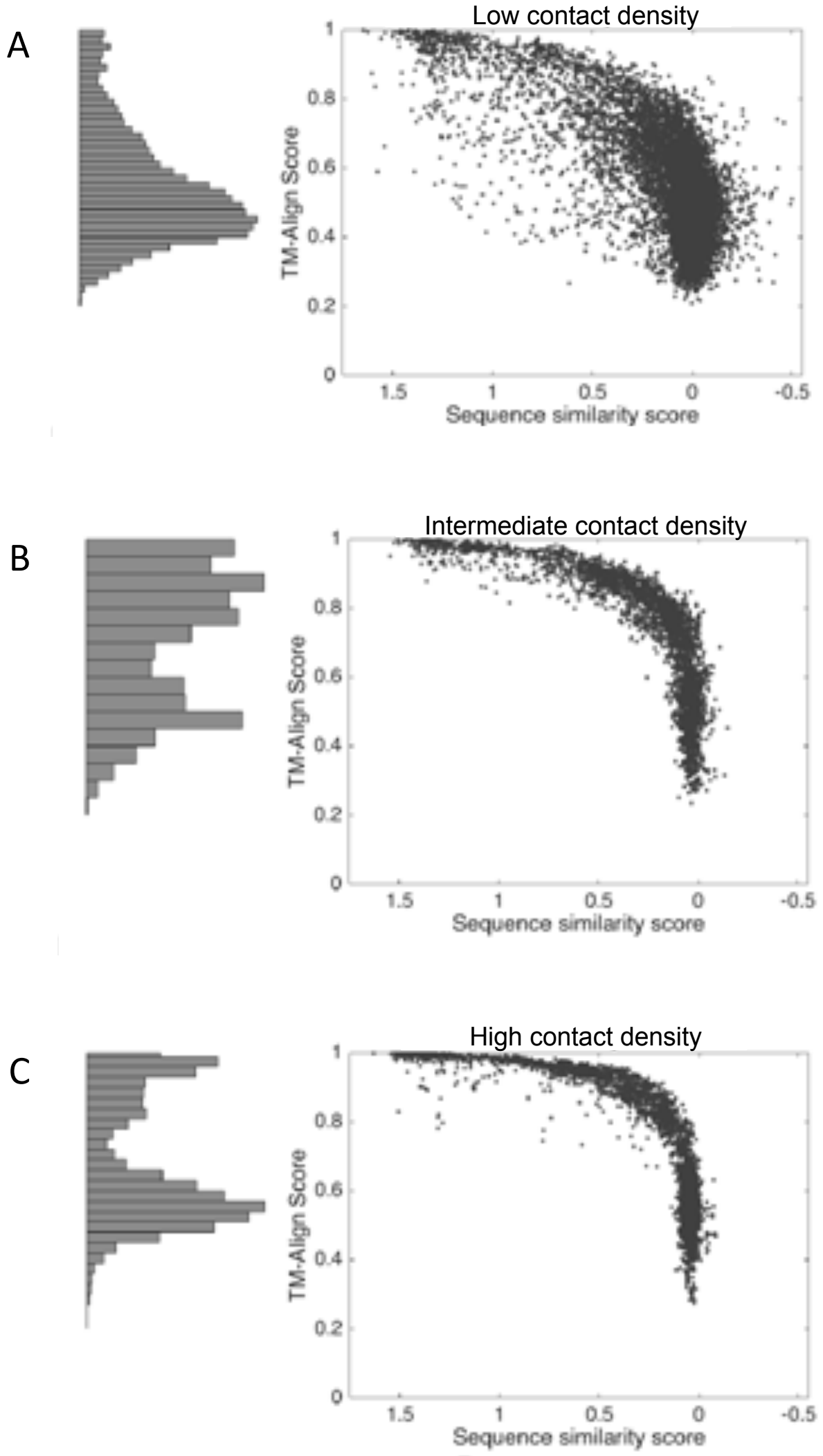

**Fig S1: Stability and fold evolution of SCOP domains** SCOP domains classified as α, β, α+β, α/β were used. The sequence similarity score (Needleman-Wunsch alignment score using the BLOSOM30 matrix) and structural homology (TM-align score) for each pair of domains classified into the same fold were calculated. (A-C) The relationship between sequence divergence and structure evolution in SCOP domains. Domains are partitioned by contact density (B) bottom 10% contact density (<4.13 contacts/residue, N=12,671 data points) (C) middle 10% contact density (4.57<CD<4.65 contacts/residue. N=3,863 data points) (D) top 10% contact density (>4.93 contacts/residue, N=5,672 data points). Histograms at the left of plots A-C are identitcal to those in Fig.1 B-D and show that at high contact density, the distribution of structure similarity TM-align scores is bimodal.

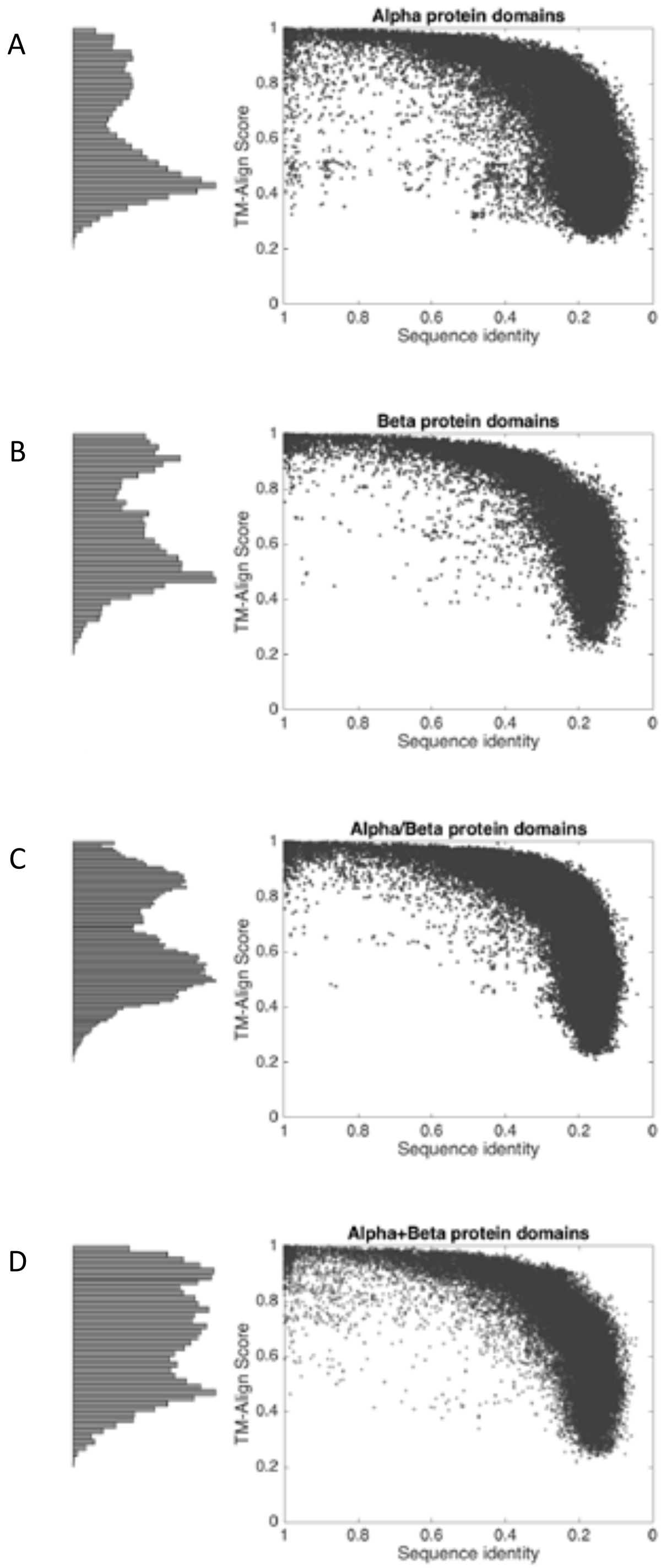

**Fig. S2. Structural divergence of α, β, α/β, and α+β protein domains from SCOP**. Sequence identity and structural homology (TM-align score) for all the domains classified into the same fold and with similar length (within 10 residues) were calculated and are plotted for (A) α (N=59,903 data points) (B) β (N=25,854 data points) (C) α/β (N=89,730 data points) and (D) α+β (N=41,264 data points) domains. Histograms at left are projections of the data onto the y-axis and are markedly bimodal across classes except, perhaps, for α+β domains. Furthermore, there is a pronounced cusped relationship between sequence and structure divergence in each class.

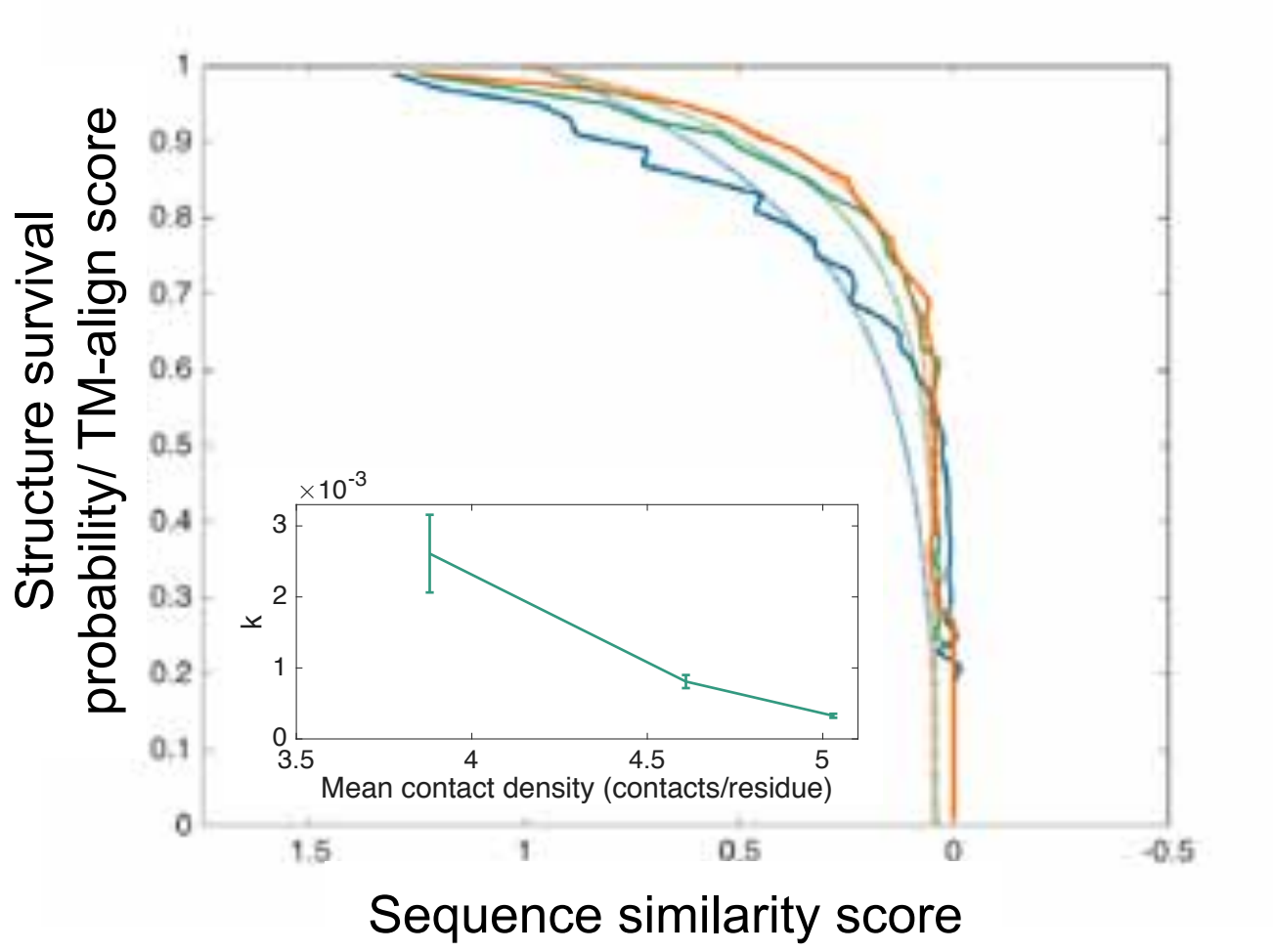

**Fig. S3.** Fit of the analytical model to the real protein data presented in Fig. S1, for which NW alignment scores were used as the measure of sequence similarity. The model fit to the data is presented in dashed lines while binned bioinformatics data is presented in solid lines for each contact density subgroup: low contact density (blue), intermediate contact density (green), high contact density (orange). Because the analytical model is based on sequence identity, it does not recover the shape of the bioinformatics data presented here as well it does when sequence identity is the metric of sequence similarity. Importantly, however, the inset shows that the negative correlation between protein domain contact density and protein evolutionary rate, $k$, predicted analytically, does not depend on the exact sequence similarity measure used. Error bars indicate one standard deviation.

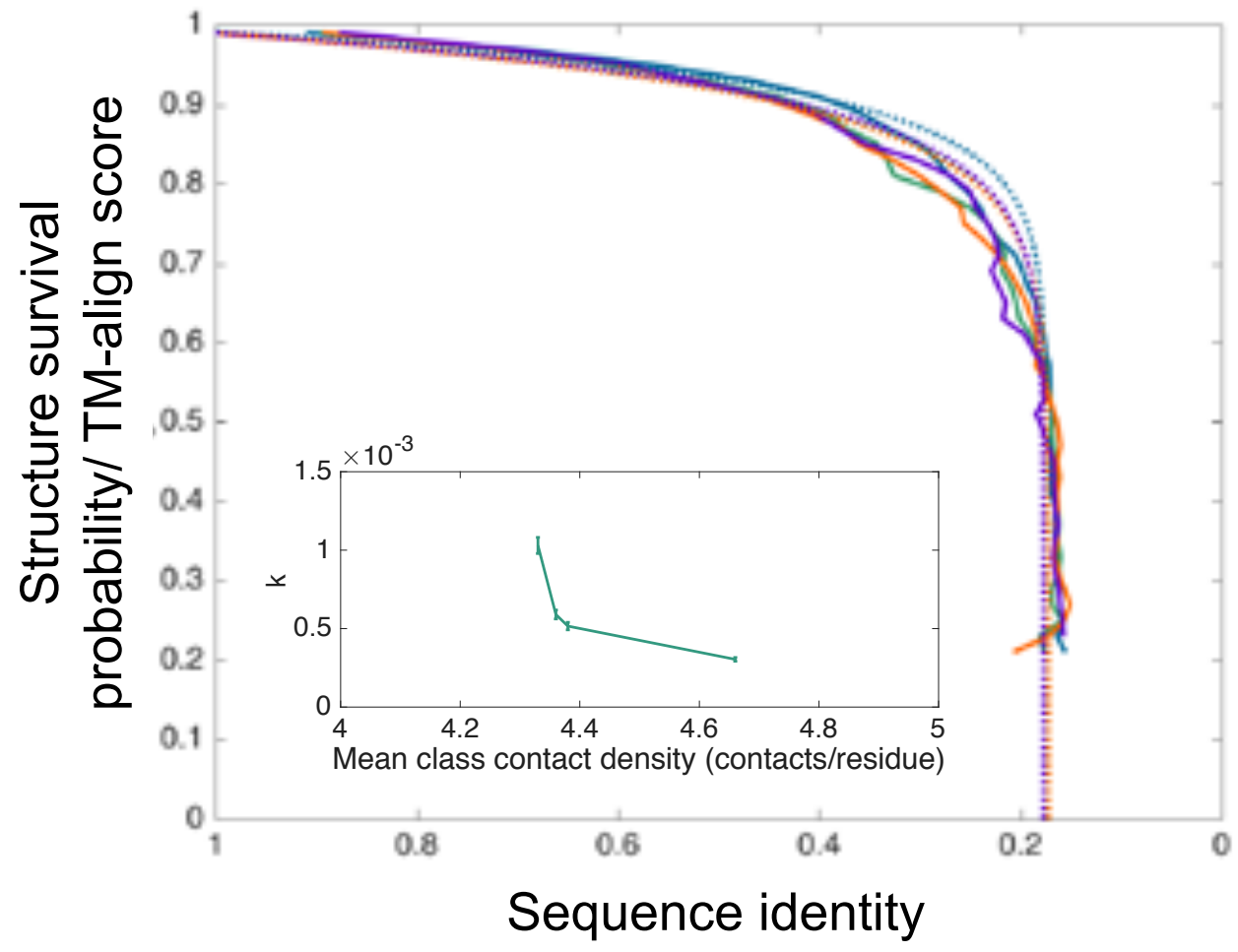

**Fig. S4.** Fit of the analytical model to data from each protein domain classes: α (purple), β (green), α/β (blue), and α+β (orange). The model fit to the data is presented in dashed lines while binned bioinformatics data is presented in solid lines. Even though there is not a clearly discernable difference among classes in the rate of structure divergence with respect to sequence divergence, the inset show that there are in fact different rates of structure evolution, $k$, across protein domains and that these differences can be explained by different mean contact densities of each class. It is unclear why the analytical fit deviates from the bioinformatics data in the cusp region here, but not when the data are decomposed by contact density. Error bars indicate one standard deviation.

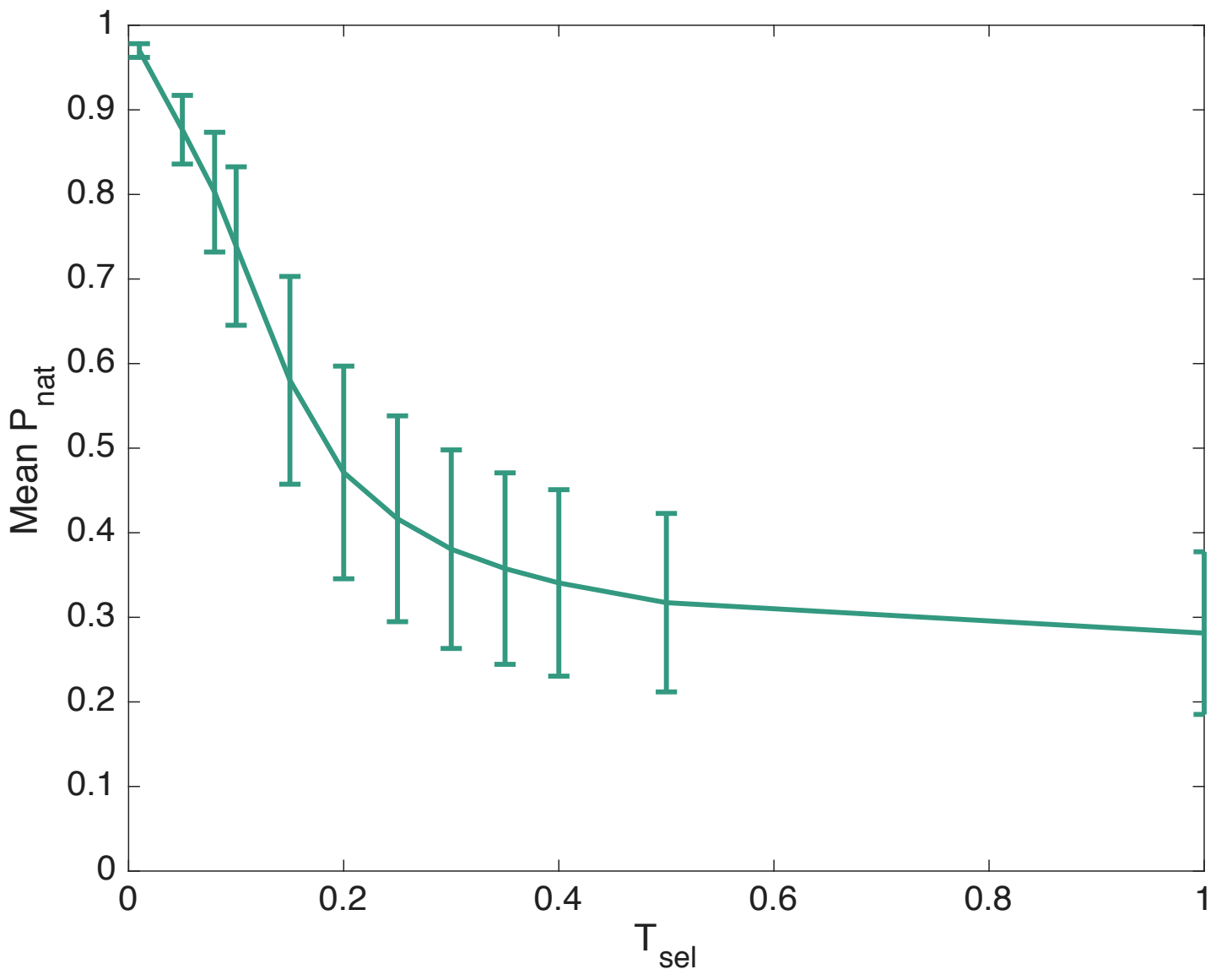

**Fig. S5.** The Monte Carlo simulation parameter, $T_{sel}$, sets the probability that a deleterious, destabilizing mutation is accepted. This adjusts the mean $P_{nat}$ at which proteins evolve. Error bars span one standard deviation in recorded $P_{nat}$ values, where the standard deviation has been averaged over runs at each $T_{sel}$. See also Dataset S2.

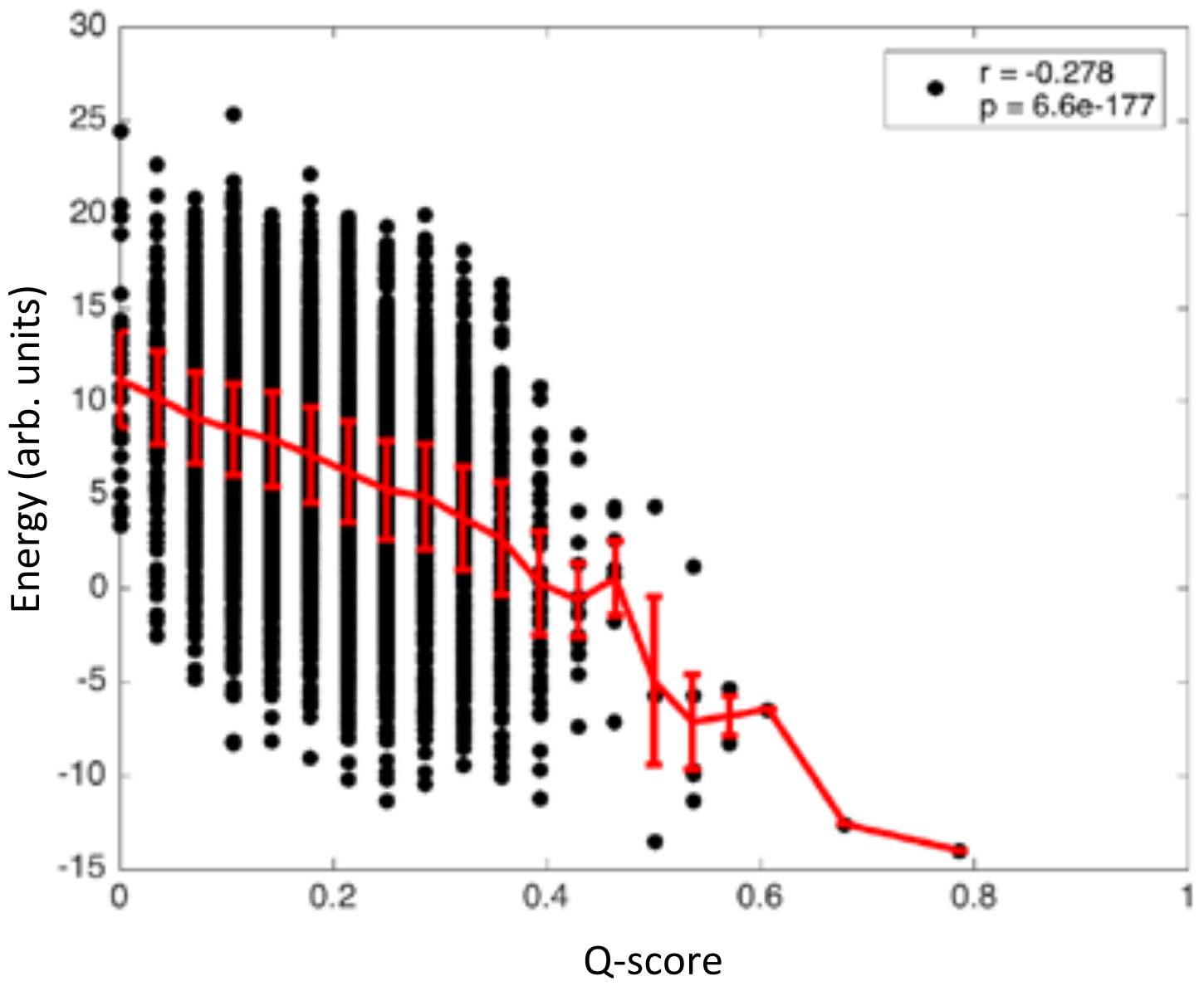

**Fig. S6.** Low-energy conformations of a sequence with a stable native structure tend to be similar (high Q-score) to the sequence's native structure. However note a broad distribution of energies with low Q-score indicating the existence of low energy misfolds that are dissimilar to the native structure.

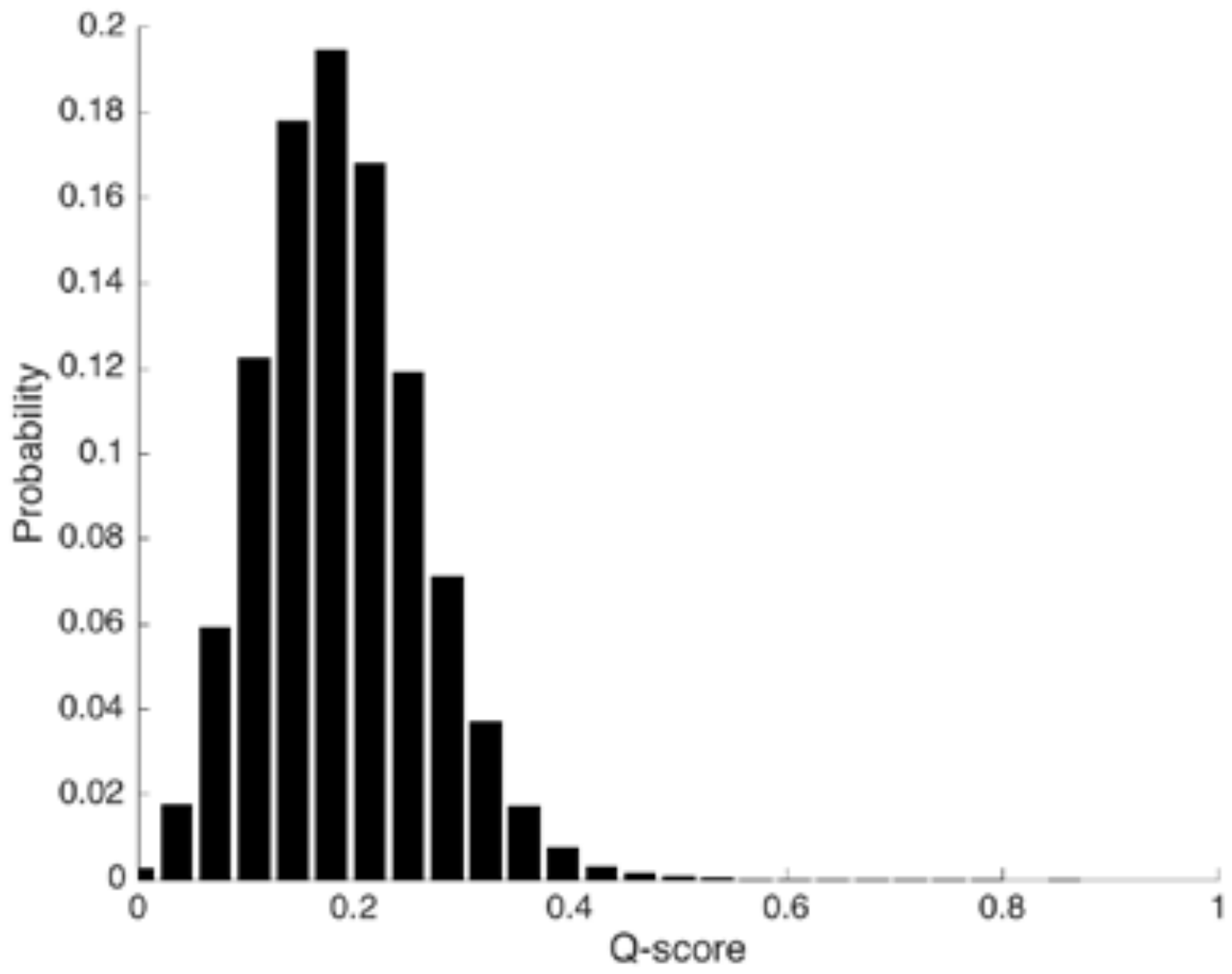

**Fig. S7.** Distribution of Q-scores for all pairs of the 10,000 structures used in this study.

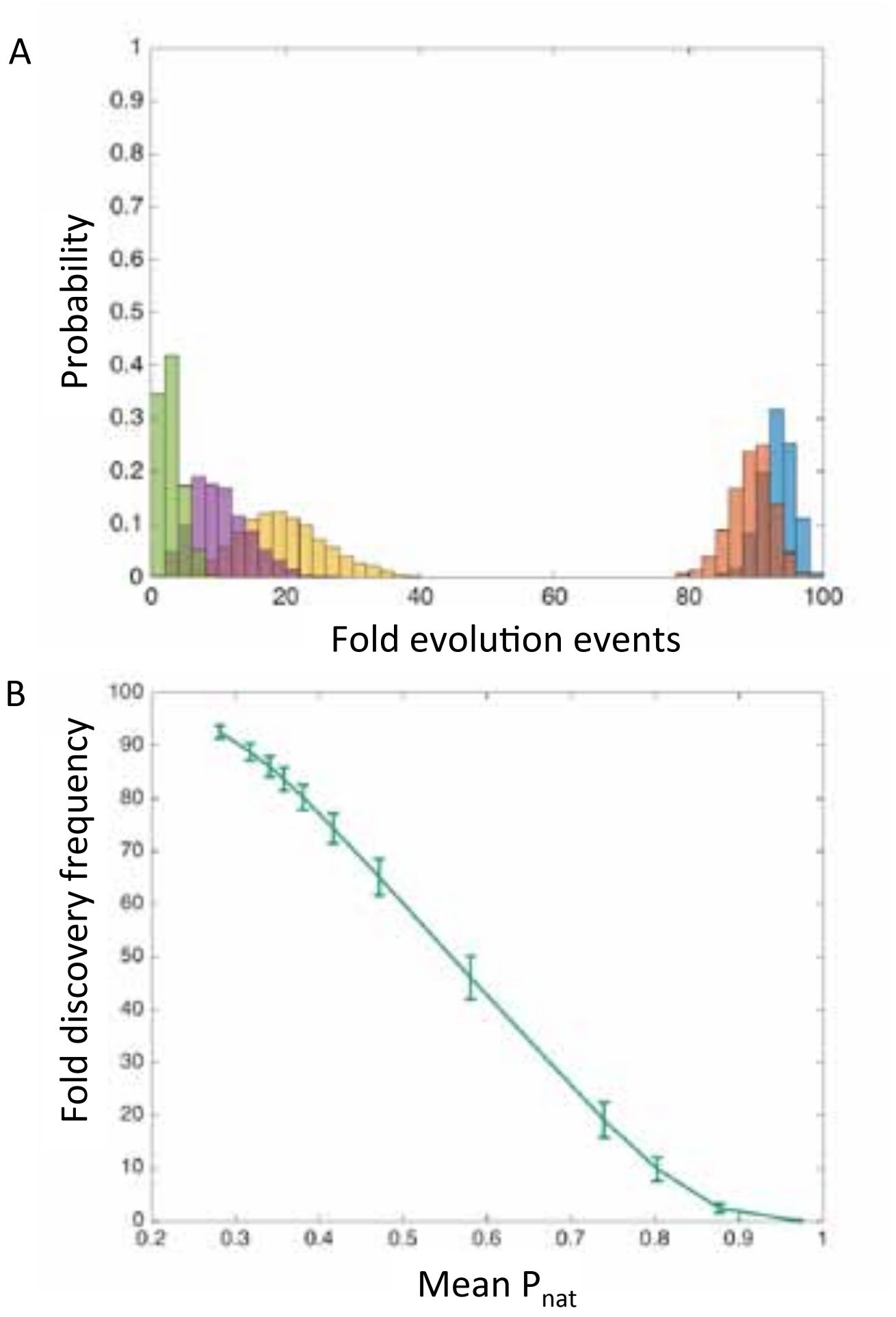

**Fig S8. Fold Discovery Events** (A) The distribution of fold evolution events over hundreds of different evolutionary runs at five different $\overline{P}_{nat}$, $\overline{P}_{nat} = 0.88$ (green), $\overline{P}_{nat} = 0.80$ (purple), $\overline{P}_{nat} = 0.74$ (yellow), $\overline{P}_{nat} = 0.32$ (orange), $\overline{P}_{nat} = 0.28$ (blue). The maximum number of fold evolution events recorded is 100 because every ten generations for 1,000 generations was recorded. (B) The number of fold discovery events averaged over evolutionary simulations run at the same selection strength. Error bars span one standard deviation.

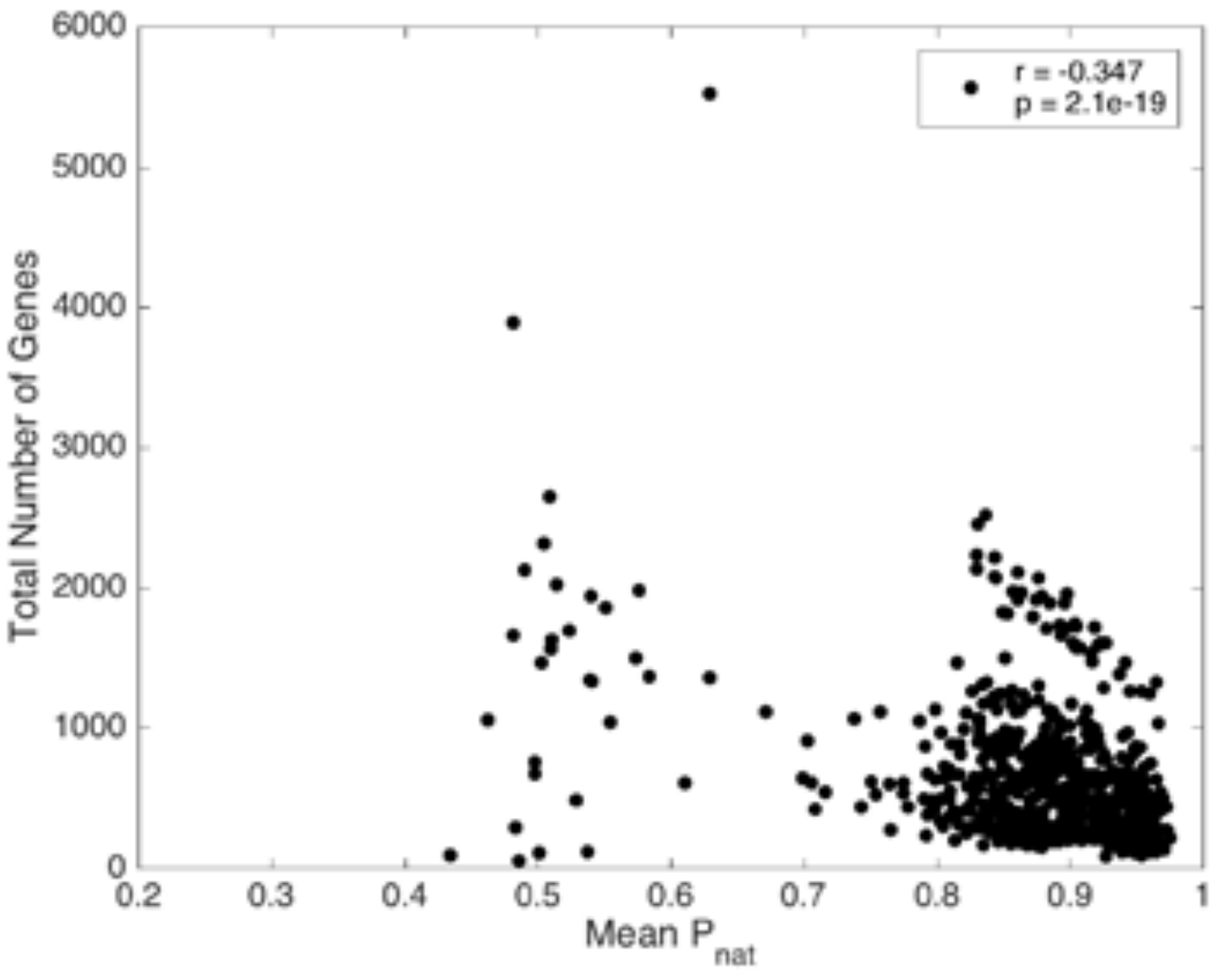

**Fig. S9.** The total number of genes in a population at the end of a population evolution simulation correlates negatively with the mean stability ($\overline{P}_{nat}$) of proteins in the population at the end of the simulation. This indicates that the positive correlation between stability and family size presented in Figure 5D is not due to differing numbers of overall proteins across simulations.

**Derivation of an analytical model of protein evolution with key gatekeeper residues**

Here, we build a phenomenological theory of protein structure evolution with additional structural, mechanistic detail. Consider a protein of length $l$ residues where each amino acid can mutate to any of the other $m = 19$ amino acids. The protein's structure is determined by $n$ key residues (1, 2). A structure evolution event occurs each time that all $n$ key residues mutate to any of $v$ amino acid types that encode a different fold, as illustrated in Fig S8A. The parameters $n$ and $v$ reflect the degree of selection pressure under which a protein is evolving. As before, the sequence evolution is captured by Eq. S9.

Rigidly fixing the number of key residues sets a hard lower bound on the minimum number of mutations before a fold evolution event can occur. Assuming that fold evolution is a Poisson process governed by mutation of key residues, we derived the "survival" probability that a fold evolution event has not occurred after time $T$ (below). With this model, both the shape and position of the twilight zone are adjusted by the strength of evolutionary selection. The requirement that all key residues must mutate before the structure evolves reflects that epistasis among the structures $n$ key residues and that extensive exploration of sequence space may be necessary before a sequence coding for a stable mutant fold is discovered.

<u>Derivation of structure survival probability time ($F(t)$)</u>
An analytic solution for the probability that the protein has undergone a structural change is derived here. For this, we will first need to calculate the characteristic wait time between fold evolution events. After $i$ "good" mutations (mutations of designated residues to any of the $v$ other amino acid types that encode another fold), the probability of a mutation being a good one is given by $P_i(good) = \frac{(n-1)v}{(l-1)m}$. The time for the protein to change structure is $T = (1/\mu)\sum_{i=1}^{n} 1/P_i$. If $H_n = \sum_{i=1}^{n} 1/i$ is the $n$ th harmonic number, then

$$T = \frac{mn}{\mu v}(1 + H_n p/n - H_n) \quad \text{(S13)}$$

Now assume that after a protein changes structure, the clock "resets." That is, after a structural change, the model is applied to the protein as if it were an ancestral state. Looking over some interval of time $\Delta t$, we expect that the protein has undergone a structural change event $\Delta t / T$ times. Therefore, after $cT$ time has elapsed, the expected number of structural changes is $c$. Now, letting the time domain be infinite, we can divide it into intervals of length $cT$, and the number of changes in each interval should be random, roughly independent, and on average $c$. The lack of independence is because there is a minimum number of time steps between structural transitions, so a transition that happens at the end of one interval will prohibit a transition at the beginning of the next. However if selection is strong, the probability distribution of the number of structural change events in an interval $cT$ should be roughly Poisson

$$P(k \text{ changes in } cT) \approx \frac{c^k}{k!} e^{-c} \tag{S14}$$

If there are $k$ changes, the average wait time $t$ between transitions is well approximated by $cT/k$, so we have that

$$P(\text{average wait time } cT/k) \approx \frac{c^{(cT/t)}}{(cT/t)!} e^{-c} \tag{S15}$$

Assume now, that $cT/t$ is continuous. If we choose $c$ such that $ce^{-c} \ll 1$, there will be only a negligible probability that $k$ is between 0 and 1, which implies that there is a negligible chance that $\tau$ is greater than $cT$. From this we can obtain a continuous cumulative distribution function (Fig S8B).

$$F(\text{structure change after time } t) = \frac{\int_0^t \frac{c^{cT/\tau}}{\Gamma(cT/\tau+1)} e^{-c} d\tau}{\int_0^{ct} \frac{c^{cT/\tau}}{\Gamma(cT/\tau+1)} e^{-c} d\tau} \tag{S16}$$

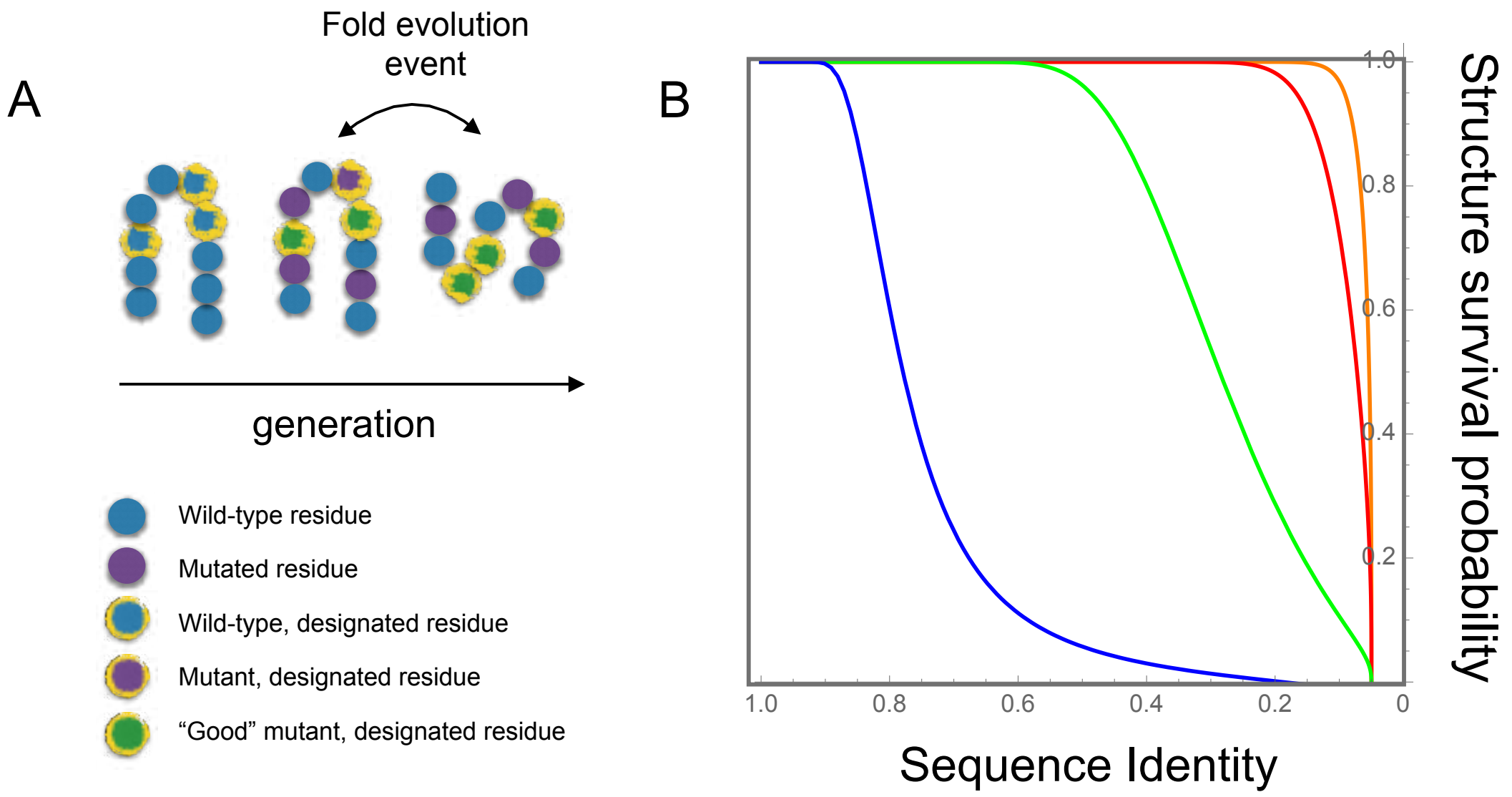

**Fig. S10. Protein evolution with key gatekeeper residues** (A) No structure or ordered sequence is explicitly modeled. Rather, in a protein of length $p$, $n$ residues are designated as residues crucial for fold stability (indicated with yellow edge). $\nu$ amino acid types out of 19 are considered "good." Once all designated residues acquire a "good" mutation (shown in green), we consider a fold discovery event to have taken place. In this model, $n$ and $\nu$ tunes the selection pressure on fold evolution. From weakest to strongest selection pressure: $n = 1$, $\nu = 19$ (blue), $n = 10$, $\nu = 10$ (green), $n = 15$, $\nu = 5$ (red), $n = 27$, $\nu = 5$ (orange).

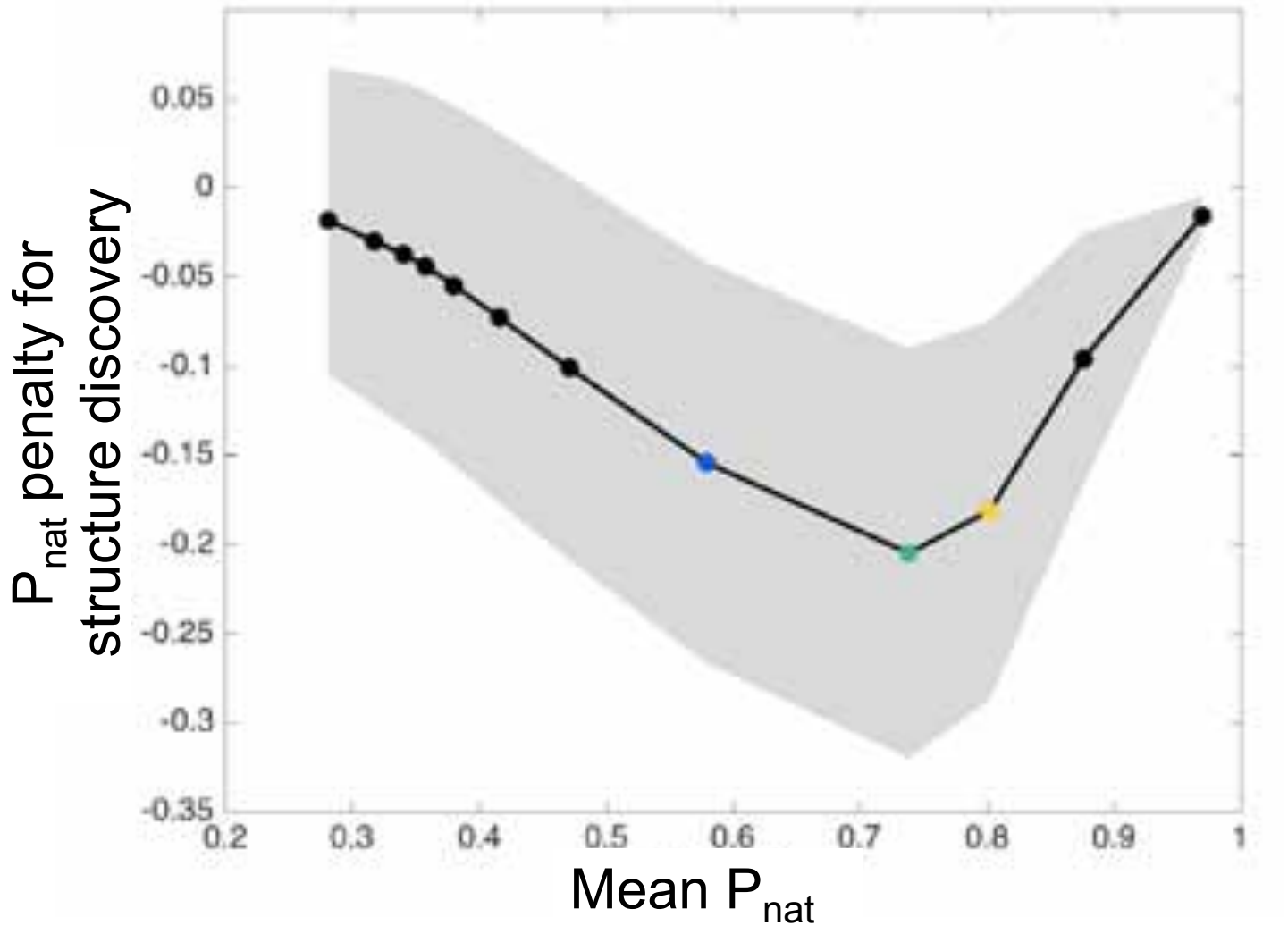

Fig. S11. Change in stability ($P_{nat}$ penalty) associated with structure evolution events for proteins evolving at different average stabilities ($\overline{P}_{nat}$). This change, shown as an average over evolutionary trajectories, is defined as the difference between protein stability the generation that a new structure emerges ($P_{nat}^{\text{(new structure)}}$) and the average stability, $\overline{P}_{nat}$, of a protein evolving at the stipulated selection temperature. The grey field spans one standard deviation. For consistency with the main text, the blue, green, and yellow points highlight $\overline{P}_{nat} = 0.58,\ 0.74,$ and $0.80$ respectively. Notably, proteins with marginal stability are *more* not less destabilized than the most stable proteins during structure discovery.

Table S1. Physical properties and analytical fitting of protein domains (sequence identity used as measure of sequence similarity)

| Protein domain category | Average domain length (number of residues) | Mean contact density (contacts/length) | k ($10^{-4}$) | c (percent sequence identity) |
|---|---|---|---|---|
| Low Contact Density (<4.13 contacts/residue) | 83.06 | 3.88 | 17.4 ± 0.7 | 0.10 ± 0.008 |
| Intermediate Contact Density (4.57<CD<4.65 contacts/residue) | 174.67 | 4.61 | 5.27 ± 0.2 | 0.12 ± 0.005 |
| High Contact Density (>4.93 contacts/residue) | 336.67 | 5.03 | 2.31 ± 0.1 | 0.12 ± 0.005 |
| α class domains | 74.96 | 4.33 | 10.33 ± 0.5 | 0.13 ± 0.007 |
| β class domains | 154.29 | 4.38 | 5.16 ± 0.2 | 0.12 ± 0.007 |
| α / β class domains | 221.29 | 4.66 | 3.04 ± 0.1 | 0.12 ± 0.005 |
| α + β class domains | 139.65 | 4.36 | 5.90 ± 0.3 | 0.12 ± 0.007 |

The average domain length and contact density of protein in each subgrouping considered along with the parameters of the analytical model fit to the datasets. The parameter $k$ is the rate of protein structure evolution and $c$ modulates the position of the twilight zone.

Table S2. Physical properties and analytical fitting of protein domains (NW-score used as measure of sequence similarity)

| Protein domain category | Average domain length (number of residues) | Mean contact density (contacts/length) | k ($10^{-4}$) | c ($10^{-3}$ x NW-score using BLOSOM30) |
|---|---|---|---|---|
| Low Contact Density (<4.13 contacts/residue) | 83.06 | 3.88 | 26.11 ± 5.5 | -18.8 ± 50 |
| Intermediate Contact Density (4.57<CD<4.65 contacts/residue) | 174.67 | 4.61 | 8.08 ± 0.9 | -9.4 ± 23 |
| High Contact Density (>4.93 contacts/residue) | 336.67 | 5.03 | 3.26 ± 0.3 | -2.1 ± 16 |

The average domain length and contact density of protein in each subgrouping considered along with the parameters of the analytical model fit to the datasets. The parameter $k$ is the rate of protein structure evolution and $c$ modulates the position of the twilight zone.

## Supporting References